\documentclass[preprint]{elsarticle}

\usepackage{amsmath}
\usepackage{amsfonts}
\usepackage{bm}
\usepackage{graphicx}
\usepackage{upgreek}

\usepackage{doi} 

\newcommand{\pdv}[2]{\frac{\partial #1}{\partial #2}}
\newcommand{\dprime}{{\prime\prime}}

\journal{Applied Mathematical Modelling}

\begin{document}
\title{A novel multiscale modelling for the hemodynamics in retinal microcirculation with an analytic solution for the capillary-tissue coupled system}

\author[label1]{Chang Lin}
\affiliation[label1]{organization={School of Mathematical Sciences, Beijing Normal University},
city={Beijing},
postcode={100875},
country={China}
}

\author[label2]{Zilong song}
\affiliation[label2]{organization={Department of Mathematics and Statistics, Utah State University},
city={Logan},
postcode={84322},
state={Utah},
country={USA}}

\author[label3]{Robert Eisenberg}
\affiliation[label3]{organization={Department of Applied Mathematics, Illinois Institute of Technology},
city={Chicago},
postcode={60616},
state={Illinois},
country={USA}}

\author[label4]{Shixin Xu}
\affiliation[label4]{organization={Zu Chongzhi Centre, Duke Kunshan University},
city={Kunshan},
country={China}
}

\author[label4]{Huaxiong Huang}

\begin{abstract}
Mathematical modelling of the microcirculatory hemodynamics in the retina is an essential tool for understanding various diseases of the retina, yet remains challenging due to the multiscale nature of the retinal vasculature and its coupling to surrounding tissue.
To address this, we develop a multiscale model that couples retinal vasculature across scales with interstitial tissue.
Our model combines the one-dimensional (1D) model for arteries and veins with the coupled Darcy equations for capillaries and tissue.
The model uses an analytic solution for capillary-tissue coupled system that provides a simple interpretation of the results along with much faster computation.
The analytic solution implies a dynamic coupling condition that links the capillary bed with upstream arterial and downstream venous flows.
The model is mathematically robust, demonstrated through analysis of the solution's truncation error and convergence.
Its predictive accuracy is verified against experimental data and other models, making it useful in interpreting experimental results.
Finally, the role of various parameters in controlling retinal hemodynamics is explored.
\end{abstract}
\maketitle

\section{Introduction}
The retina is a thin and multilayered neural tissue lining the back of the eye that plays a crucial role in visual function.
Its function depends sensitively on adequate blood supply, as retinal hemodynamics helps sustain the metabolism that keeps the retina alive and functional \cite{kornfield_measurement_2015}.
Disruptions in the hemodynamics can impair oxygen and nutrient delivery, contributing to various retinopathies such as diabetic retinopathy (DR) \cite{kornfield_measurement_2015}, glaucoma \cite{garhofer_response_2004,caprioli_blood_2010,chng_ocular_2021} and retinal vascular occlusion \cite{osborne_retinal_2004}.
Understanding retinal hemodynamics is thus essential for understanding the life of the retina in health and disease.

Computational modelling serves as a crucial and powerful tool for exploring retinal hemodynamics, integrating physiological processes that are difficult to measure \textit{in vivo} \cite{ruffini_mathematical_2024,siggers_fluid_2012}.
Different models have been employed in retinal heomodynamics and oxygenation.
Lumped parameter zero-dimensional (0D) models, representing the vasculature as hydraulic resistor circuit \cite{flores_novel_2016}, have helped to resolve the inconsistencies in clinical data \cite{guidoboni_intraocular_2014}.
The 0D accumulated model, in which each segment is modelled as a resistor, is exploited in the simulation of retinal circulation \cite{ganesan_analysis_2010,fry_predicting_2018}.
More complicated models, including one-dimensional (1D) model, two-dimensional (2D) model and three-dimensional (3D) model, have been used to reveal fundamental properties of oxygen transport in retina \cite{liu_computational_2009,causin_blood_2016} as well as deal with properties of disease in specific patients \cite{julien_one-dimensional_2023,rebhan_computational_2019,malek_computational_2015,tripathy_image-based_2023}.
Existing models provide valuable physiological and pathological insights, however, they focus on specific vascular compartments, including simplifications such as constant capillary pressure or neglecting interactions between capillaries and surrounding tissue.
Consequently, very few existing models combine the arterioles, venules, capillaries and tissue equally into a comprehensive model.
In other tissues, the general rule that all compartments interact with each other has required a systematic structural analysis of flow \cite{eisenberg_structural_2023}, and the same should hold true for the retina.

A central challenge in modelling the hemodynamics of retinal microcirculation stems from intrinsic multiscale vasculature.
Arteriolar and venular networks display hierarchical tree-like branching \cite{julien_one-dimensional_2023}, while capillaries form an interconnected mesh-like structure \cite{fry_predicting_2018,shipley_hybrid_2019}.
Thus retinal microcirculation inherently constitutes a multiscale system, involving small scale details of flow inside blood vessels and also the large scale layout of the blood vessels.
These issues can be addressed using existing methodologies for other organs or tissues, including tumour perfusion \cite{desposito_computational_2018}, transport in breast tissue \cite{fritz_1d0d3d_2022}, and microcirculatory hemodynamics \cite{sweeney_threedimensional_2024,shipley_hybrid_2019} where a continuum Darcy model for capillary flow is typically coupled to arterial and venous trees via point sources.
Even with the point source approximation, the computation is  challenging because the Darcy model has to be solved simultaneously with arterial and venous flows \cite{qohar_nonlinear_2021,fritz_1d0d3d_2022}.
A novel alternative is to use the analytic solution for the capillary flow, which expresses the pressure explicitly so that the Darcy model does not have to be solved in the entire domain. The analytical solution provides significant physical insight as well as much faster computation \cite{shipley_hybrid_2019,sweeney_threedimensional_2024,desposito_computational_2018}.
However in current analytic methods the fluid exchange between capillaries and tissue is treated approximately.
They describe the coupling of capillaries to surrounding tissues using a far-field boundary condition.
While this serves as a reasonable first step, it does not capture the local fluid exchange between these two components, and a more realistic description should include such exchange.

Our work complements previous studies by developing a novel multiscale model that fully couples retinal vasculature with interstitial tissue.
Specifically, the flows in arterioles and venules are described by a 1D model \cite{julien_one-dimensional_2023}, while capillaries and surrounding tissue are treated as two coupled porous media and the flow in them is described by the Darcy equation.
The fluid exchange between the capillary and tissue is modelled by the source/sink of the two Darcy equations, one for capillary (drainage) and the other for the tissue (feed).
Moreover, the vasculature is coupled as follows.
Terminal arterioles/venules in superficial layer act as point sources/sinks in the capillary Darcy's equation, and the pressure difference between these vessels and the capillary network is related to the flow rates at point sources/sinks by the effective resistors in deep layer, which provides the outlet/inlet boundary condition for arteriolar/venular tree in superficial layer.
What further distinguishes our modelling from others is the analytic solution for the capillary-tissue system, which is derived on a finite domain with physiological boundary conditions.
We analyse the convergence and truncation error of the analytical solution extensively to ensure its robustness and accuracy.
The analysis of the solution's convergence also reveals some interesting properties, which might suggest a new modelling methodology for other tissues. 
The model is validated against experimental data and other results from existing models, and pulsatile pressures within arterioles and venules are also simulated.
Moreover we demonstrate our model's capability by exploring the effects of parameter variations.

The rest of the paper is organized as follows.
In Section$~$\ref{Models} we describe the models for vasculature and surrounding tissue, including the 1D model for arterioles and venules, synthetic vascular network model, the continuum formulation for the capillary-tissue system.
In Section$~$\ref{AnalyticSolutionCapillaryTissueCoupledSystem} we derive the analytic solution of capillary-tissue coupled system, and its convergence is analysed.
In Section$~$\ref{NumericaMethods} the numerical methods for proposed model are described, and multiple simulations are performed to assess the model's accuracy and demonstrate its capability in Section$~$\ref{ValidationProposedMultiscaleModel}.
The sensitivity of various parameters is explored in Section$~$\ref{SensitivityAnalysis}.
Section$~$\ref{DiscussionAndConclusion} consists of the conclusion and discussion.

\section{Models}
\label{Models}
\subsection{Network-based simulation for arteriolar and venular trees}
\subsubsection{One-dimensional model}
The one-dimensional (1D) model is widely adopted to describe the blood flow in the individual vessels of  arteriolar and venular networks. The networks exist in three dimensional space but are often analysed in a two dimensional subset of three dimensional space following the usual conventions of electric circuit theory. For all time $t>0$ the flow within each vessel is characterized by the flow rate $q(s,t)$ and the cross-sectional area $A(s,t)$ \cite{julien_one-dimensional_2023, wang_fluid_2016}, along the axis $s$ of the vessel, and their evolutions are given by
\begin{flalign}
    \left\{
    \begin{aligned}
        &\pdv{A}{t}+\pdv{q}{s}=0, \\
        &\pdv{q}{t}+\pdv{}{s}\left(\hat{\alpha}\frac{q^2}{A}\right)+\frac{A}{\rho}\pdv{p}{s}=C_f,
        \label{1D_model}
    \end{aligned}\right.
\end{flalign}
where $\hat{\alpha}$ is the Boussinesq coefficient (momentum flux correction factor) assumed to describe the velocity profile in each vessel \cite{muller_global_2014}, $p$ is the pressure, $\rho$ is the density of blood and $C_f$ is the friction term.
The coefficient $\hat{\alpha}$ is set as $\hat{\alpha}=4/3$ which corresponds to the parabolic velocity profile that is consistent with the experimental data in the retina \cite{song_combined_2014,causin_blood_2016,kornfield_measurement_2015}.
The density is set as $\rho=1~{\rm{g}/\rm{cm}^3}$ from now on in this paper.
The explicit form of friction term is given by \cite{muller_global_2014}
\begin{eqnarray}
    C_f=\left\{\begin{aligned}
        &-8\pi\frac{\eta}{\rho}\frac{q}{A},\quad \text{for arterioles},\\
        &-8\pi\sqrt{\frac{A}{A_0}}\frac{\eta}{\rho}\frac{q}{A},\quad \text{for venules},
    \end{aligned}\right.
\end{eqnarray}
where $\eta=\eta(r)$ is the dynamical viscosity modelled as a function of the vessel radius $r$ to account for the rheological properties of blood.

The fluid-structure interaction in the problem is captured by the following elastic model \cite{liu_multicore_2024, muller_global_2014}:
\begin{eqnarray}
    p-p_{ext}=K\left[\left(\frac{A}{A_0}\right)^{\beta_1}-\left(\frac{A}{A_0}\right)^{\beta_2}\right],
    \label{elastic_model}
\end{eqnarray}
where $p_{ext}$ is the exterior pressure, $K$ is the rigidity being the intrinsic feature of vessel, $\beta_1$ and $\beta_2$ are parameters depending on the vessel wall behaviour, and $A_0$ is the neutral area of vessel, which differs for each vessel.
The values of $K$, $\beta_1$ and $\beta_2$ vary with the vessel position and its function.
The rigidity is
\begin{eqnarray}
    K=\left\{\begin{aligned}
        &\frac{4}{3}\frac{\pi^{1/2}Eh}{A^{1/2}_0}, \quad \text{for arterioles},\\
        &\frac{1}{9}\frac{\pi^{3/2}Eh^3}{A^{3/2}_0}, \quad \text{for venules}.
    \end{aligned}\right.
\end{eqnarray}
with $E$ the Young modulus and $h$ the thickness of the vessel wall.
The parameters $\alpha$ and $\beta$ are given by
\begin{eqnarray}
    \beta_1=\left\{\begin{aligned}
        &1/2,\quad \text{for arterioles},\\
        &10,\quad \text{for venules},
    \end{aligned}\right.
\end{eqnarray}
and
\begin{eqnarray}
    \beta_2=\left\{\begin{aligned}
        &0,\quad \text{for arterioles},\\
        &-3/2,\quad \text{for venules}.
    \end{aligned}\right.
\end{eqnarray}
Expressions of $K$ and values of $\beta_1$ and $\beta_2$ for arterioles are mainly taken from previous work on modelling of arteriolar circulation \cite{julien_one-dimensional_2023}, while those for venules are adopted from the model of veins in other tissues \cite{muller_global_2014,audebert_kinetic_2017}.
The values of $E$ for arterioles and venules are set to the same value as a reasonable approximation \cite{julien_one-dimensional_2023}.
The wall thickness $h$ is proportional to the vessel diameter, which is given by $h=\rm{WLR}\times 2r_0$ where $\rm{WLR}$ is the wall-to-lumen ratio and $r_0$ is the radius of neutral area $A_0$. For arterioles, this value is based on experimental observations \cite{arichika_effects_2015}, while for venules, a smaller value is employed to account for their thinner walls.
Values of parameters in 1D model are summarised in table$~$\ref{parameters_retinal_circulation}.

The apparent viscosity of blood decreases strongly with the decreasing vessel radius from about $150~\rm{\upmu m}$ to about $5~\rm{\upmu m}$, then increases below $5~\rm{\upmu m}$.
This unique rheological behaviour is the Fåhræus–Lindqvist effect.
To characterize blood's rheological properties, we use an empirical relation developed by Pries \textit{et al.} that expresses viscosity as a function of vessel radius \cite{pries_resistance_1994,secomb_blood_2013,bappoo_viscosity_2017}:
\begin{eqnarray}
\eta(r)=\eta_p\left[1+(\eta_{45}-1)\left(\frac{r}{r-0.55}\right)^2\right]\left(\frac{r}{r-0.55}\right)^2,
\end{eqnarray}
where $r$ is the vessel radius (in unit $\rm{\upmu m}$), $\eta_p$ is the viscosity of plasma set as $\eta_p=0.012~\rm{Poise}$ and $\eta_{45}$ is the relative apparent blood viscosity for discharge hematocrit of $0.45$ given by
\begin{eqnarray}
    \eta_{45}=6\exp(-0.17r)+3.2-2.44\exp\left[-0.06(2r)^{0.645}\right].
\end{eqnarray}

Note that the 1D model is adopted for superficial vessels, which are those visible in fundus images.
As a result, the external pressure acting on the vessel walls is approximately the intraocular pressure (IOP) \cite{kiel_ocular_2011}, thus in the 1D model, external pressure is set as $p_{ext}=\rm{IOP}$.
Similar approximation is employed in previous work \cite{guidoboni_intraocular_2014}.
Smaller vessels are distributed in the deep layer and invisible in fundus images, which makes them difficult to distinguish clinically.
The standard method is to model these deep vessels as effective resistors that connect the superficial vascular trees to the capillary bed.
The synthetic vascular network model for both superficial and deep vessels is described below.
\subsubsection{Synthetic vascular network model}
\begin{figure}
	\centering
	\includegraphics[width=\textwidth]{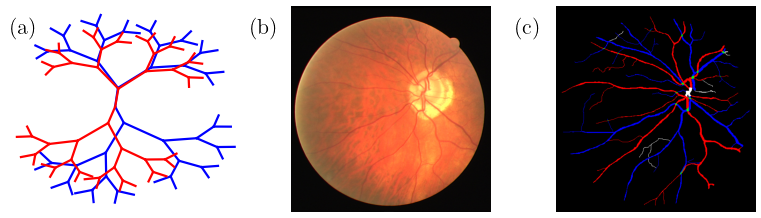}
	\caption{Synthetic and real retinal vasculature. (a) The arteriolar (red) and venular (blue) trees in retinal tissue (black circle) generated using an L-system. (b) Fundus image from DRIVE dataset \cite{staal_ridge-based_2004}. (c) The manually segmented retinal vasculature for (b). Arterioles and venules are displayed in red and blue, respectively. The green denotes pixels classified as both arteriole and venule and the white denotes the uncertain segment.}
	\label{L_system_arterial_venous_trees}
\end{figure}

Next we construct a network from individual vessels.
We exploit a Lindenmayer system (L-system) to generate synthetic arteriolar and venular trees in superficial layer \cite{brown_physics-informed_2024}.
The vasculature begins with the central retinal artery (CRA) and central retinal vein (CRV), which are perpendicular to the tissue at the center of the optic disc.
Branches are added iteratively and asymmetrically to create bifurcations up to the fifth generation, with lengths given by $l_v=\xi r_v$, where $\xi$ is the length to radius ratio and $r_v$ is the vessel radius.
At each bifurcation boundary one mother vessel is connected to two daughter vessels, and two branching angles $\theta_{d_1}$ and $\theta_{d_2}$ are determined by \cite{brown_physics-informed_2024,zamir_arterial_2001}
\begin{eqnarray}
    \cos\theta_{d_1}=\frac{(1+v_a^3)^{4/3}+v_a^4-1}{2v_a^2(1+v_a^3)^{2/3}},\\
    \cos\theta_{d_2}=\frac{(1+v_a^3)^{4/3}+1-v^4_a}{2v_a^2(1+v_a^3)^{2/3}},
\end{eqnarray}
where $v_a=r_{d_2}/r_{d_1}<1$ is the asymmetry ratio, with radii $r_{d_1}$ and $r_{d_2}$ of two daughter vessels, and normally-distributed noise is added to branching angle values with a standard deviation of $\pi/36$.
The vessel radii follow asymmetric branching relation
\begin{eqnarray}
    r^\gamma_M=r^\gamma_{d_1}+r^\gamma_{d_2},
\end{eqnarray}
where $r_M$ is the radius of mother vessel and $\gamma$ is the junction exponent \cite{brown_physics-informed_2024,zamir_arterial_2001}.
Given the radius of root vessel $r_M$ and the value of $v_a$, radii of two daughter vessels are determined by \cite{zamir_arterial_2001,liu_computational_2009}
\begin{eqnarray}
    r_{d_1}=(1+v^\gamma_a)^{-1/\gamma}r_M,\label{daughter_one_vessel_radius} \\ r_{d_2}=v_a(1+v^\gamma_a)^{-1/\gamma}r_M.\label{daughte_two_vessel_radius}
\end{eqnarray}
Generated vascular trees are shown in figure$~$\ref{L_system_arterial_venous_trees}(a), where the arteriolar tree is displayed in red and venular tree is displayed in blue.
For comparison, a real retinal vasculature from the DRIVE dataset \cite{staal_ridge-based_2004} and its manual segmentation are shown in figure$~$\ref{L_system_arterial_venous_trees}(b) and (c), respectively.

Arterioles and venules in deep layer are constructed using a structured tree model to set the resistive boundary condition for the vascular trees in superficial layer \cite{liu_computational_2009,malek_computational_2015,julien_one-dimensional_2023}.
This boundary condition is defined by the Poiseuille equation for steady, viscous flow, which describes the relation between flow and pressure, given by
\begin{eqnarray}
    p_{up}-p_{down}=Rq,
    \label{Poiseuill_equation_structured_tree}
\end{eqnarray}
where $ p_{up}$ and $p_{down}$ are the upstream and downstream pressures, $R$ is the resistance and $q$ is flow rate.

In the structured tree model deep vessels would grow from terminal superficial vessels until their radii are decreased to $r=6~\rm{\upmu m}$, which is the characteristic radius of capillary \cite{jozsa_porous_2021,fritz_1d0d3d_2022}, and the branching is performed similarly as that in L-system \cite{julien_one-dimensional_2023}.
The radius of vessel $r_v$ is computed using equations$~$(\ref{daughter_one_vessel_radius}) or (\ref{daughte_two_vessel_radius}), while the length is given by $l_v=\xi r_v$.
Given the radius and length, the resistance of a single vessel is given by
\begin{eqnarray}
    R_k=\frac{8\eta_k l_k}{\pi r^4_k},
    \label{resistance_arteriole_venule}
\end{eqnarray}
where the viscosity $\eta_k$ is computed by Pries \textit{et al.}'s model.
Then the equivalent resistance of the root vessel in a structured tree is calculated recursively for resistive boundary condition.
For a vessel $M$ with its own resistance $R_M$ and its daughter vessels with resistances $R_{d_1,eq}$ and $R_{d_2,eq}$, the total equivalent resistance is
\begin{eqnarray}
    R_{M,eq}=R_M+R_{d,eq},
    \label{equivalent_mother_vessel_resistance}
\end{eqnarray}
where
\begin{eqnarray}
    \frac{1}{R_{d,eq}}=\frac{1}{R_{d_1,eq}}+\frac{1}{R_{d_2,eq}}.
    \label{equivalent_resistance_three_branch}
\end{eqnarray}
\subsection{Continuum formulation for the capillary network and surrounding tissue}
\subsubsection{Physiological basis and modelling simplification}
In our model we treat the retina as a porous medium, given its composition of cells, fibres and interstitial space filled with a fluid.
The capillary bed forms multiple interconnected plexuses, which are superficial, intermediate, and deep plexuses, rather than a collection of a few discrete, isolated branches.
These plexuses are densely distributed in the neural tissue, and they are compressed into a single porous medium such that the essential hemodynamics could be effectively described.
Given bidirectional fluid exchange between capillaries and surrounding tissue, these two systems should be combined in the same way that electrical systems are combined in electrical models of syncytial tissue \cite{eisenberg_structural_2023}.
We thus model the capillary bed and surrounding tissue as interacting porous media.
\subsubsection{Capillary-tissue coupled system}
The flow in both capillary bed and surrounding tissue is modelled as a Darcy flow \cite{sweeney_threedimensional_2024,shipley_hybrid_2019,qohar_nonlinear_2021,fritz_1d0d3d_2022}.
The Darcy flux of capillaries ${\bm u}_{cap}$, i.e., volumetric flow rate per unit area (in unit ${\rm{cm}/ \rm{s}}$), is given by
\begin{eqnarray}
    {\bm u}_{cap}=-\frac{1}{\mu_{cap}}{\bm k}_{cap}\cdot\nabla p_{cap}\quad \text{in $\Omega_t$},
    \label{Darcy_flow}
\end{eqnarray}
where ${\bm k}_{cap}$ is the permeability tensor (in unit ${\rm{\upmu m}^2}$), $\mu_{cap}$ is the viscosity of blood and $p_{cap}$ is the pressure of capillary network.
The equation$~$(\ref{Darcy_flow}) states that the flow in capillaries is mainly driven by the pressure gradient, which is consistent with flow characteristics in retinal vasculature physiologically.
The capillary network is assumed isotropic and homogeneous \cite{qohar_nonlinear_2021,vidotto_hybrid_2019} such that the permeability tensor is ${\bm k}_{cap}=k_{cap}{\bm I}$, where ${\bm I}$ is the identity matrix.
Since the blood is incompressible conservation of mass yields
\begin{eqnarray}
    \nabla\cdot{\bm u}_{cap}=\psi_{cap}\quad \text{in $\Omega_t$},
    \label{Darcy_mass_conservation}
\end{eqnarray}
where $\psi_{cap}$ is the source term (in unit ${\rm{s}^{-1}}$).
Combining equations$~$(\ref{Darcy_flow}) and (\ref{Darcy_mass_conservation}) the flow in capillaries is characterized by following equation
\begin{eqnarray}
    -\frac{k_{cap}}{\mu_{cap}}\nabla^2 p_{cap}=\psi_{cap}({\bm x} )\quad \text{in $\Omega_t$}.
\end{eqnarray}
The source term should combine three key components: arteriolar inflow, venular drainage and the fluid exchange between capillaries and surrounding tissue, which is expressed as
\begin{eqnarray}
    \psi_{cap}({\bm x})=\psi_{ac}+\psi_{cv}-\psi_{ct},
    \label{capillary_source}
\end{eqnarray}
where $\psi_{ac}$, $\psi_{cv}$ and $\psi_{ct}$ denote arteriolar inflow, venular drainage and fluid exchange between capillaries and the tissue, respectively.
The contributions of arteriolar and venular trees are modelled as the summation of discrete point sources:
\begin{eqnarray}
    \psi_{ac}({\bm x})+\psi_{cv}({\bm x})=\sum^{N_{cap}}_{n=1}q_n\delta({\bm x}-{\bm x}_n),
\end{eqnarray}
where $N_{cap}$ is the total number of point sources, $q_n$ represents the flow rate of source/sink at location ${\bm x}_n$ and $\delta(\cdot)$ is the Dirac delta function.
The fluid exchange between capillaries and tissue is mainly determined by their hydrostatic and osmotic pressures.
Using the Starling's filtration principle the fluid filtration rate is given by \cite{xu_osmosis_2018,fritz_1d0d3d_2022,vidotto_hybrid_2019}
\begin{eqnarray}
    \psi_{ct}({\bm x})=L_h\times S\times [(p_{cap}({\bm x})-p_t({\bm x}))-\alpha_r(\pi_{cap}({\bm x})-\pi_t({\bm x}))],
\end{eqnarray}
where $L_h$ is the hydraulic conductivity of capillary wall, $S$ is the capillary surface area per unit volume of tissue, $p_t$ is the tissue pressure in retinal layers, $\pi_{cap}$ is capillary plasma osmotic pressure, $\pi_t$ is interstitial fluid osmotic pressure and $\alpha_r$ is the permeability of capillary wall to proteins.
Due to the tight junctions of the blood-retina barrier (BRB), the capillary wall is largely impermeable to plasma proteins, which implies that $\alpha_r\approx 1$ under normal physiological conditions \cite{truskey_transport_2009}. For our model focusing on the hemodynamics, the osmotic pressure difference $\Delta \pi=\pi_{cap}-\pi_t$ is treated as a constant, which is reasonable when the protein leakage is minimal and capillary protein concentrations do not vary dramatically along the vascular network, consistent with our continuum-scale averaging over capillaries.
The source $\psi_{ct}$ could be written as
\begin{eqnarray}
    \psi_{ct}({\bm x})=\alpha_{exch}\times (p_{cap}({\bm x})-p_t({\bm x})-\alpha_r\Delta\pi),
\end{eqnarray}
where $\alpha_{exch}=L_h\times S$ is the fluid exchange rate (in unit ${\rm{\upmu m}\cdot\rm{s}/\rm{g}}$).
Similarly the tissue pressure satisfies
\begin{eqnarray}
    -\frac{k_t}{\mu_t}\nabla^2 p_t=\psi_{ct}, \quad \text{in $\Omega_t$},
\end{eqnarray}
where $k_t$ is the permeability for the interstitial fluid in tissue, $\mu_t$ is the viscosity of interstitial fluid in tissue.
Note that active pumping of retinal pigment epithelium (RPE) and passive exchange with the vitreous chamber and choroid \cite{ruffini_mathematical_2024,dvoriashyna_osmotic_2018} are not included, since these mechanisms mainly regulate long-term fluid homeostasis and baseline pressure, rather than the local pressure differences that drive capillary-tissue exchange.
Then the coupled system for capillary and interstitial fluid pressure is given by
\begin{eqnarray}
    \left\{
    \begin{aligned}
        &-\frac{k_{cap}}{\mu_{cap}}\nabla^2 p_{cap}=\sum^{N_{cap}}_{n=1}q_n\delta({\bm x}-{\bm x}_n)-\alpha_{exch}(p_{cap}-p_t-\alpha_r\Delta\pi) \quad \text{in $\Omega_t$},\\
        &-\frac{k_t}{\mu_t}\nabla^2 p_t=\alpha_{exch}(p_{cap}-p_t-\alpha_r\Delta\pi) \quad \text{in $\Omega_t$}.
    \end{aligned}\right.
    \label{capillary_tissue_coupled_model}
\end{eqnarray}
Previous studies usually assume an idealized geometry of the tissue domain $\Omega_t$, such as a sphere \cite{siggers_fluid_2012,rebhan_computational_2019,tripathy_image-based_2023}, a circular domain with radial cuts \cite{fry_predicting_2018,ganesan_analysis_2010}, or an equivalent rectangle with the same area \cite{causin_blood_2016}, which is motivated by the spherical anatomy of the retina.
Following this precedent, we assume a circular domain of radius $R_t$ for the tissue.

\subsubsection{Boundary conditions}
We next specify the boundary conditions for capillary and tissue pressures.
Physiologically, blood should not flow across the boundary $\partial\Omega_t$.
Since the flux of blood across the boundary is described by $-(k_{cap}/\mu_{cap})\nabla p_{cap}\cdot {\bm n}$, and enforcing this flux to be zero implies following boundary condition for $p_{cap}$:
\begin{eqnarray}
    -\frac{k_{cap}}{\mu_{cap}}\nabla p_{cap}\cdot {\bm n}=0 \quad \text{on $\partial\Omega_t$}.
\end{eqnarray}
For the tissue pressure, since all fluid exchanges are included in source the flux of interstitial fluid on boundary is zero, and thus the boundary condition for tissue pressure is given by
\begin{eqnarray}
   -\frac{k_t}{\mu_t}\nabla p_t\cdot {\bm n}=0 \quad \text{on $\partial\Omega_t$}.
\end{eqnarray}
Moreover these two boundary conditions correspond to two fundamental conservation laws.
First, integrating the boundary condition for $p_t$ over $\partial\Omega_t$ yields
\begin{eqnarray}
    \int_{\partial\Omega_t}-\frac{k_t}{\mu_t}\nabla p_t\cdot {\bm n}ds=\int_{\Omega_t}-\frac{k_t}{\mu_t}\nabla^2 p_td{\bm x}=\int_{\Omega_t}\alpha_{exch}(p_{cap}-p_t-\alpha_r\Delta\pi)d{\bm x}=0,
\end{eqnarray}
which implies that
\begin{eqnarray}
    \int_{\Omega_t}p_{cap}d{\bm x}=\int_{\Omega_t}p_td{\bm x}+\frac{\alpha_r}{\alpha_{exch}}\int_{\Omega_t}\Delta\pi d{\bm x}.
    \label{pressure_filtration_p_cap_p_t}
\end{eqnarray}
This equality states the tissue fluid homeostasis, where the integrated capillary and interstitial fluid pressures are balanced by the oncotic pressure difference, ensuring that no pathological fluid accumulation occurs in the interstitium.
Second, integrating the boundary condition for $p_{cap}$ implies
\begin{eqnarray}
    \int_{\partial\Omega_t}-\frac{k_{cap}}{\mu_{cap}}\nabla p_{cap}\cdot {\bm n}ds&=&\int_{\Omega_t}\psi_{cap}d{\bm x} \nonumber\\
    &=&\sum^{N_{cap}}_{n=1}q_n-\int_{\Omega_t}\alpha_{exch}(p_{cap}-p_t-\alpha_r\Delta\pi)d{\bm x},
\end{eqnarray}
thus the boundary condition that $-(k_{cap}/\mu_{cap})\nabla p_{cap}\cdot {\bm n}=0$ is satisfied only if $\sum^{N_{cap}}_{n=1}q_n=0$, which states the conservation of mass.
\subsection{Boundary conditions for vascular tree and multiscale coupling}
It remains to specify the boundary conditions for the 1D model, as well as the coupling conditions at the interfaces where arterioles and venules connect to the capillaries.

\begin{figure}
	\centering
	\includegraphics[width=\textwidth]{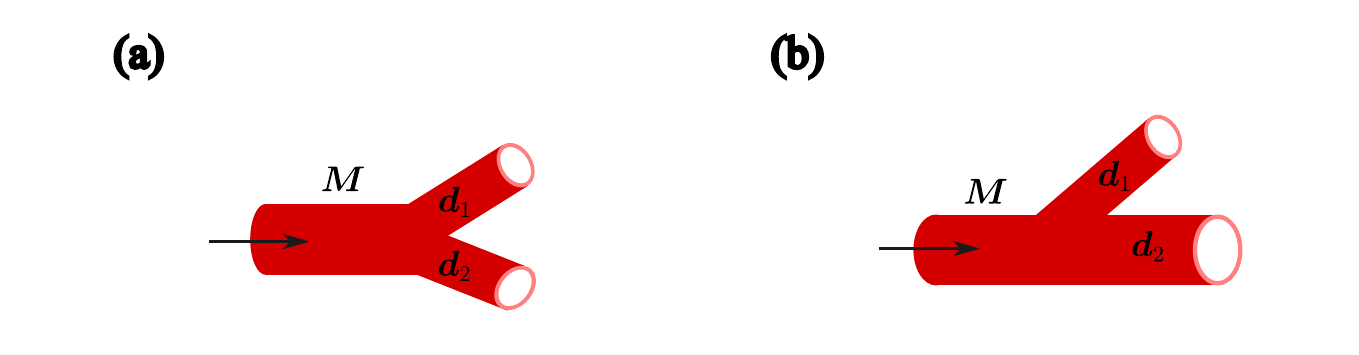}
	\caption{Two different bifurcation boundaries.}
	\label{bifurct_boundary}
\end{figure}

The pressure boundary conditions are imposed at the CRA inlet and CRV outlet.
While the pressure at CRA could be constant or time-dependent, that at CRV outlet is set constant.
We denote areas at CRA inlet and CRV outlet as $A_{in,CRA}$ and $A_{out,CRV}$, respectively.
The area $A_{in,CRA}$ is determined by solving the equation
\begin{eqnarray}
   p_{in,CRA}-p_{ext}=K\left[\left(\frac{A_{in,CRA}}{A_0}\right)^{\beta_1}-\left(\frac{A_{in,CRA}}{A_0}\right)^{\beta_2}\right],
\end{eqnarray}
where $p_{in,CRA}$ is the prescribed pressure at CRA inlet, and the corresponding flow rate $q_{in,CRA}$ is obtained from the mass-conservation equation
\begin{eqnarray}
    \pdv{A_{CRA}}{t}+\pdv{q_{CRA}}{s}=0.
\end{eqnarray}
Similarly, the area $A_{out,CRV}$ is computed from
\begin{eqnarray}
    p_{out,CRV}-p_{ext}=K\left[\left(\frac{A_{out,CRV}}{A_0}\right)^{\beta_1}-\left(\frac{A_{out,CRV}}{A_0}\right)^{\beta_2}\right],
\end{eqnarray}
where $p_{out,CRV}$ is the prescribed pressure at CRV outlet.
At CRV outlet, constant pressure implies a constant area $A_{out,CRV}$, and the mass-conservation equation then yields the boundary condition that $\partial q/\partial s=0$.
The bifurcation boundary conditions are governed by following equations \cite{koppl_dimension_2023}:
\begin{eqnarray}
    \left\{
    \begin{aligned}
        &q_M-q_{d_1}-q_{d_2}=0,\\
        &p_M-p_{d_1}=0,\\
        &p_M-p_{d_2}=0,\\
        &\pdv{A_M}{t}+\pdv{q_M}{s_M}=0,\\
        &\pdv{A_{d_1}}{t}+\pdv{q_{d_1}}{s_{d_1}}=0,\\
        &\pdv{A_{d_2}}{t}+\pdv{q_{d_2}}{s_{d_2}}=0,
    \end{aligned}\right.
\end{eqnarray}
where the subscripts $M$, $d_1$ and $d_2$ denote the mother and two daughter vessels, respectively.
These conditions are identical for different shapes of bifurcation boundary, as shown in figure$~$\ref{bifurct_boundary}.

The multiscale coupling between the vascular tree and capillary-tissue system is imposed at source points ${\bm x}={\bm x}_n$ with $n=1,\dots,N_{cap}$.
In this work, source points are uniformly distributed to ensure full perfusion of the tissue.
These points correspond to the positions where structured trees connect the superficial vessels to capillary bed, and the flow within each structured tree is described by the Poiseuille equation$~$(\ref{Poiseuill_equation_structured_tree}).
Specifically, the pressure-flow relations for structured trees connecting to superficial arterioles and venules are given by
\begin{eqnarray}
    p_{out,arteriole}-p_{cap}({\bm x}_n)=R_{a,eq}({\bm x}_n)q_n, \label{pressure_flow_arteriole}\\
    p_{cap}({\bm x}_i)-p_{in,venule}=R_{v,eq}({\bm x}_i)q_i, \label{pressure_flow_venule}
\end{eqnarray}
respectively, where $R_{a,eq}({\bm x}_n)$ and $R_{v,eq}({\bm x}_i)$ are the equivalent resistances of structured trees at ${\bm x}_n$ and ${\bm x}_i$, respectively, computed using equations$~$(\ref{equivalent_mother_vessel_resistance}), (\ref{equivalent_resistance_three_branch}), and (\ref{resistance_arteriole_venule}).
The variables $p_{out,arteriole}$ and $p_{in,venule}$ are outlet and inlet pressures of the the terminal vessel in the vascular tree, respectively, and $q_n$ is the flow rate at terminal arteriole outlet (or venule inlet).
The coupling conditions$~$(\ref{pressure_flow_arteriole}) and (\ref{pressure_flow_venule}) are imposed using the analytic solution for $p_{cap}$, which is derived in the following section.

\subsection{Summary of multiscale model}
\begin{figure}
    \centering
    \includegraphics[width=\textwidth]{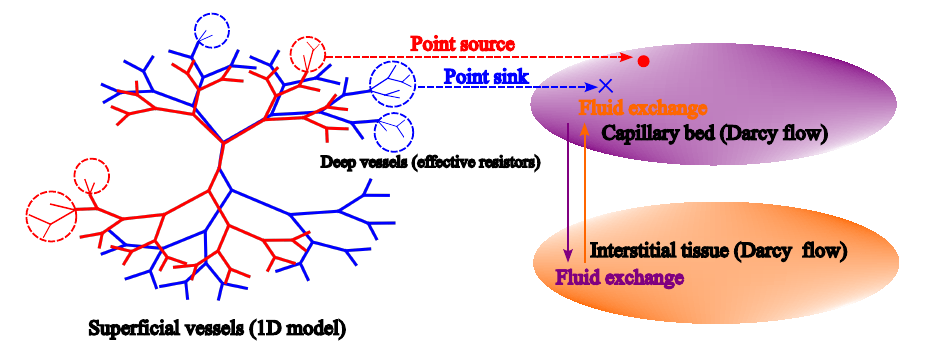}
    \caption{Schematic of the multiscale model. The arteriolar and venular trees in the superficial layer and those in the deep layer (vessels within dashed circles) are shown on the left, while the capillary bed and interstitial tissue are depicted on the right. The flow in superficial vessels is described by the 1D model, whereas the flow in capillary bed and interstitial tissue is modelled as the Darcy flow. The vessels in deep layer are modelled as effective resistors to set the resistive boundary condition for the arteriolar/venular trees in superficial layer. The vasculature is coupled through the point sources (red circle) and sinks (blue cross) located in capillary bed, and capillary continuum and interstitial tissue is coupled through fluid exchange, denoted by orange and purple arrows.}
    \label{arteries_veins_capillaries_tissue}
\end{figure}

Arterioles and venules in the superficial layer of retina are constructed using an L-system, and the flow in them is described by the 1D model$~$(\ref{1D_model}).
Smaller vessels in the deep layer are generated using a structured tree model and treated as effective resistors, providing resistive boundary conditions for the 1D model.
The superficial arterioles and venules are coupled to capillaries through discrete point sources, with the coupling impose via the pressure-flow relations$~$(\ref{pressure_flow_arteriole}) and (\ref{pressure_flow_venule}).
The capillary bed and surrounding tissue are modelled as interacting porous media, described by two coupled continuum Darcy models$~$(\ref{capillary_tissue_coupled_model}).
This multiscale model enables a comprehensive description of retinal hemodynamics, from large feeding vessels to the capillary level, including the fluid exchange between the vasculature and the interstitial tissue.
A schematic of the multiscale model is shown in figure$~$\ref{arteries_veins_capillaries_tissue}.

\section{Analytic solution for capillary-tissue coupled system}
\label{AnalyticSolutionCapillaryTissueCoupledSystem}
\subsection{Solutions for capillary and tissue pressures}
We denote the capillary and tissue conductivities as $\kappa_{cap}=k_{cap}/\mu_{cap}$ and $\kappa_t=k_t/\mu_t$, respectively.
With transformed capillary pressure $\tilde{p}_{cap}=p_{cap}-\alpha_r\Delta\pi$ the coupled system can be expressed as
\begin{eqnarray}
    \left\{
    \begin{aligned}
        &-\kappa_{cap}\nabla^2 \tilde{p}_{cap}=\sum^{N_{cap}}_{n=1}q_n\delta({\bm x}-{\bm x}_n)-\alpha_{exch}(\tilde{p}_{cap}-p_t) \quad \text{in $\Omega_t$},\\
        &-\kappa_t\nabla^2 p_t=\alpha_{exch}(\tilde{p}_{cap}-p_t)+\psi_{water} \quad \text{in $\Omega_t$},
    \end{aligned}\right.
\end{eqnarray}
which is subject to following boundary conditions
\begin{eqnarray}
    \left\{
        \begin{aligned}
            &-\kappa_{cap}\nabla \tilde{p}_{cap}\cdot {\bm n}=0 \quad \text{on $\partial\Omega_t$}, \\
            &-\kappa_t\nabla p_t\cdot {\bm n}=0 \quad \text{on $\partial\Omega_t$}.
        \end{aligned}\right.
    \label{capillary_tissue_coupled_system_bc}
\end{eqnarray}
To solve the capillary-tissue coupled system we begin with writing the equations in matrix form
\begin{eqnarray}
    -\nabla^2\begin{bmatrix}
        \tilde{p}_{cap} \\
        p_t
    \end{bmatrix}
    ={\bm C}\begin{bmatrix}
        \tilde{p}_{cap} \\
        p_t
    \end{bmatrix}+\begin{bmatrix}
        \frac{1}{\kappa_{cap}}\sum_nq_n\delta({\bm x}-{\bm x}_n) \\
        0
    \end{bmatrix},
\end{eqnarray}
where the matrix ${\bm C}$ is given by
\begin{eqnarray}
    {\bm C}=\alpha_{exch}\begin{bmatrix}
        -1/\kappa_{cap} & 1/\kappa_{cap} \\
        1/\kappa_t & -1/\kappa_t
    \end{bmatrix}.
\end{eqnarray}

The system is decoupled via the following transformation to new variables:

\begin{eqnarray}
    \begin{bmatrix}
        p_{mean} \\
        p_{exch}
    \end{bmatrix}={\bm T}
    \begin{bmatrix}
        \tilde{p}_{cap} \\
        p_t
    \end{bmatrix},
\end{eqnarray}
where $p_{mean}$ is the conductivity weighted average pressure, $p_{exch}$ is the exchange pressure proportional to their difference, and ${\bm T}$ is the transformation matrix given by
\begin{eqnarray}
    {\bm T}=\frac{\kappa_{cap}}{\kappa_{cap}+\kappa_t}
    \begin{bmatrix}
        1 & \kappa_t/\kappa_{cap} \\
        -1 & 1
    \end{bmatrix}.
\end{eqnarray}
Using this transformation, the boundary conditions become
\begin{eqnarray}
    \left\{
    \begin{aligned}
        &-\kappa_{cap}p_{mean}\cdot {\bm n}=0 \quad \text{on $\partial\Omega_t$}, \\
        &-\kappa_t\nabla p_{exch}\cdot {\bm n}=0 \quad \text{on $\partial\Omega_t$}.
    \end{aligned}\right.
\end{eqnarray}
The resulting decoupled system is then given by
\begin{eqnarray}
    \left\{
    \begin{aligned}
        &-\nabla^2p_{mean}=\frac{1}{\kappa_{cap}+\kappa_t}\sum^{N_{cap}}_{n=1}q_n\delta({\bm x}-{\bm x}_n) \quad \text{in $\Omega_t$},\\
        & -\nabla^2p_{exch}+\lambda^2p_{exch}=-\frac{1}{\kappa_{cap}+\kappa_t}\sum^{N_{cap}}_{n=1}q_n\delta({\bm x}-{\bm x}_n)\quad \text{in $\Omega_t$},
    \end{aligned}\right.
\end{eqnarray}
where $\lambda=\sqrt{\alpha_{exch}(1/\kappa_{cap}+1/\kappa_t)}$ and the system is subject to the boundary conditions
\begin{eqnarray}
    \left\{
    \begin{aligned}
        &\nabla p_{mean}\cdot {\bm n}=0 \quad \text{on $\partial\Omega_t$}, \\
        &\nabla p_{exch}\cdot {\bm n}=0 \quad \text{on $\partial\Omega_t$}.
    \end{aligned}\right.
\end{eqnarray}
Once the decoupled system are solved the transformed capillary pressure and tissue pressures can be obtained through the inverse transformation that

\begin{eqnarray}
    \begin{bmatrix}
        \tilde{p}_{cap} \\
        p_t
    \end{bmatrix}={\bm T}^{-1}\begin{bmatrix}
        p_{mean} \\
        p_{exch}
    \end{bmatrix}=\begin{bmatrix}
        p_{mean}-(\kappa_t/\kappa_{cap})p_{exch} \\
        p_{mean}+p_{exch}
    \end{bmatrix},
\end{eqnarray}

where ${\bm T}^{-1}$ is the inverse transformation matrix given by

\begin{eqnarray}
	{\bm T}^{-1}=\begin{bmatrix}
        1 & -\kappa_t/\kappa_{cap} \\
        1 & 1
    \end{bmatrix}.
\end{eqnarray}
The capillary pressure could be obtained by adding the osmotic pressure difference to transformed capillary pressure: 
\begin{eqnarray}
    p_{cap}=\tilde{p}_{cap}+\alpha_r\Delta\pi.
\end{eqnarray}

\begin{figure}
    \centering
    \includegraphics[width=0.47\textwidth]{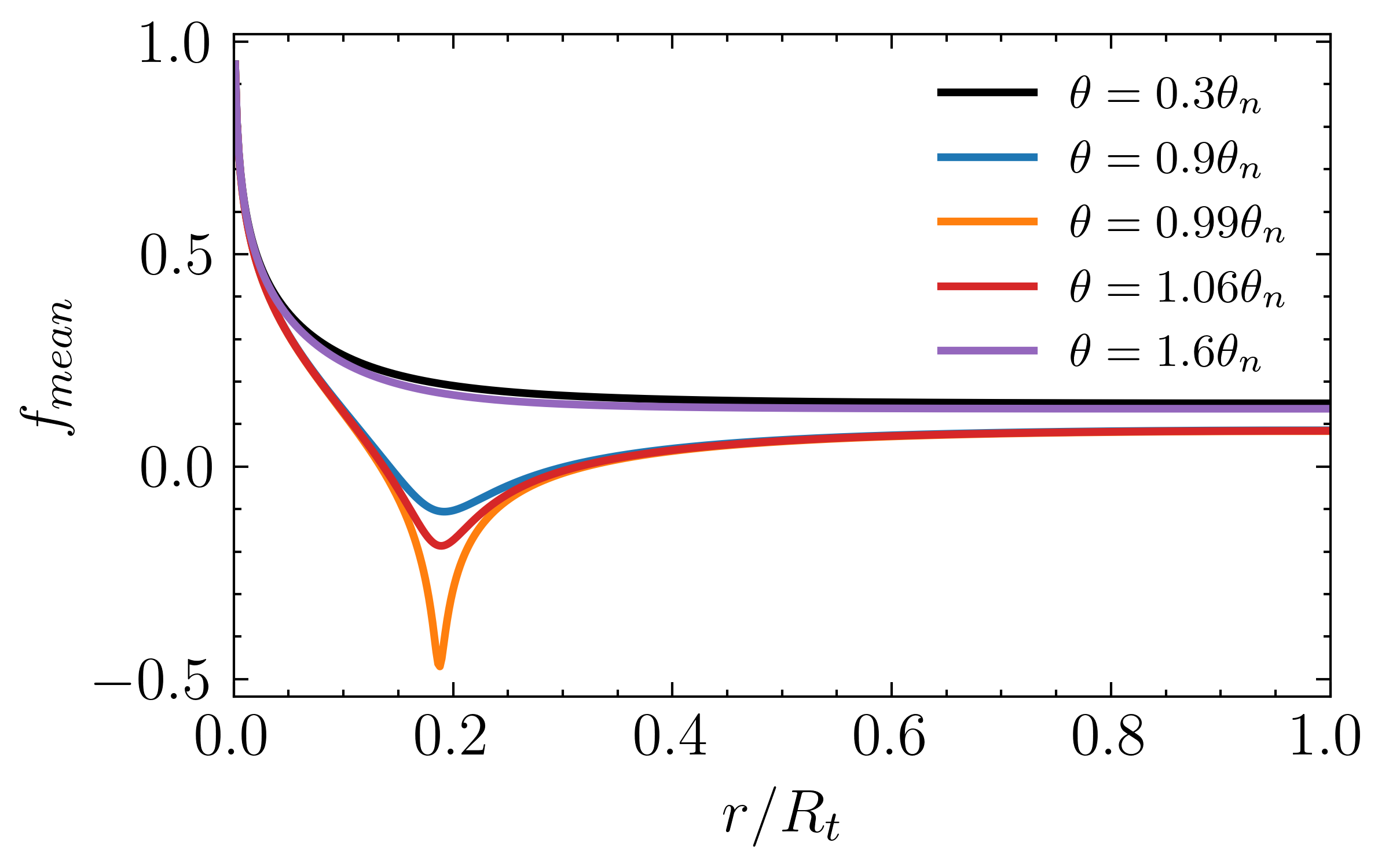}
    \caption{Behaviour of $f_{mean}$ on $r/R_t$ for a single source at $(r_n,\theta_n)=(0.19R_t,2\pi/3)$. Different colours correspond to various values of $\theta$.}
    \label{f_mean_variation}
\end{figure}

To solve the decoupled system, we construct the relevant Green's functions.
The Green's function $f_{mean}({\bm x},{\bm x}^\prime)$ for $p_{mean}$ satisfies \cite{kim_proof_1993,riley_mathematical_2006}
\begin{flalign}
    \left\{
    \begin{aligned}
        &{\nabla^\prime}^2f_{mean}({\bm x},{\bm x}^\prime)=\delta({\bm x}-{\bm x}^\prime)\quad &\text{in $\Omega_t$}, \\
        &\nabla^\prime f_{mean}({\bm x},{\bm x}^\prime)\cdot {\bm n}^\prime=1/S\quad &\text{on $\partial\Omega_t$},
        \label{p_mean_unit_source}
    \end{aligned}\right.
\end{flalign}
where $S=2\pi R_t$ is the length of the boundary $\partial\Omega_t$.
To construct the solution for mean pressure, an image point source is added at ${\bm x}^\dprime\in \mathbb{R}^2\setminus \Omega_t$, where ${\bm x}^\dprime=(R^2_t/\vert {\bm x}\vert^2){\bm x}$.
The general solution for $f_{mean}$ is then given by
\begin{eqnarray}
    f_{mean}({\bm x},{\bm x}^\prime)=\frac{1}{2\pi}\left(\log\vert{\bm x}^\prime-{\bm x}\vert+k\log \vert{\bm x}^\prime-{\bm x}^\dprime\vert+c\right),
\end{eqnarray}
where $k$ and $c$ are the constants left to be determined.
The normal derivative is computed as
\begin{eqnarray}
    \nabla^\prime f_{mean}({\bm x},{\bm x}^\prime)\cdot {\bm n}^\prime&=&\frac{1}{2\pi}\left[\frac{{\bm x}^\prime-{\bm x}}{\vert{\bm x}^\prime-{\bm x}\vert^2}+\frac{k({\bm x}^\prime-{\bm x}^\dprime)}{\vert{\bm x}^\prime-{\bm x}^\dprime\vert^2}\right]\cdot\frac{{\bm x}^\prime}{\vert {\bm x}^\prime\vert}\nonumber\\
    &=&\frac{1}{2\pi \vert {\bm x}^\prime\vert}\left[\frac{\vert {\bm x}^\prime\vert^2-{\bm x}\cdot{\bm x}^\prime}{\vert{\bm x}^\prime-{\bm x}\vert^2}+\frac{k\vert {\bm x}^\prime\vert^2-k{\bm x}^\prime\cdot{\bm x}^\dprime}{\vert{\bm x}^\prime-{\bm x}^\dprime\vert^2}\right],
\end{eqnarray}
and on the boundary $\partial \Omega_t$, where $\vert {\bm x}^\prime\vert=R_t$, this expression simplifies to
\begin{eqnarray}
    \nabla^\prime f_{mean}({\bm x},{\bm x}^\prime)\cdot {\bm n}^\prime=\frac{1}{2\pi R_t}\frac{1}{\vert{\bm x}^\prime-{\bm x}\vert^2}\left[\vert {\bm x}^\prime\vert^2+k\vert {\bm x}\vert^2-(k+1){\bm x}^\prime\cdot{\bm x}\right].
\end{eqnarray}
We can obtain the Green's function  $f_{mean}$ and the mean pressure by enforcing the boundary condition in problem$~$(\ref{p_mean_unit_source}) to solve for $k$.
The computation of coefficients in $f_{mean}$ and the derivation for $p_{mean}$ are provided in Appendix \ref{SolutionMeanPressure}.
The mean pressure $p_{mean}$ is given by
\begin{eqnarray}
    p_{mean}({\bm x})=-\frac{1}{\kappa_{cap}+\kappa_t}\sum^{N_{cap}}_{n=1}q_nf_{mean}({\bm x},{\bm x}_n)+\overline{p_{mean}},
\end{eqnarray}
where the boundary average $\overline{p_{mean}}$ is given by
\begin{eqnarray}
    \overline{p_{mean}}=\frac{1}{S}\int_{\partial\Omega_t}p_{mean}({\bm x}^\prime) ds^\prime.
\end{eqnarray}
We write the variable $p_{mean}$ in polar coordinates as
\begin{eqnarray}
    p_{mean}(r,\theta)=-\frac{1}{\kappa_{cap}+\kappa_t}\sum^{N_{cap}}_{n=1}q_nf_{mean}(r,\theta;r_n,\theta_n)+\overline{p_{mean}},
    \label{decoupled_pressure_p_mean}
\end{eqnarray}
where $r_n$ and $\theta_n$ are radial and polar coordinates for source point ${\bm x}_n$, respectively, and the function $f_{mean}(r,\theta;r_n,\theta_n)$ is given by
\begin{eqnarray}
   f_{mean}(r,\theta;r_n,\theta_n)&=&\frac{1}{4\pi}\log(r^2_n+r^2-2r_nr\cos(\theta_n-\theta)) \nonumber\\
    &&+\frac{1}{4\pi}\log\left(r^2_n+\frac{R^4_t}{r^2}-2r_n\frac{R^2_t}{r}\cos(\theta_n-\theta)\right).
    \label{f_mean_solution}
\end{eqnarray}
Figure$~$\ref{f_mean_variation} displays $f_{mean}$ as a function of $r/R_t$ for $r_n=0.19R_t$ and $\theta_n=2\pi/3$.
Different colours correspond to different values of $\theta$.
All curves diverge as $r\to 0$, and when $\theta$ is close to $\theta_n$, $f_{mean}$ develops an increasingly sharp peak at $r=r_n$, as shown by blue, orange, and red curves, which is the second logarithmic singularity.
When $r\to R_t$, $f_{mean}$ tends finite values that depend on $\theta$.

When implementing the solution for $p_{mean}$, the conservation of mass should be enforced explicitly, which is given by
\begin{eqnarray}
    0=\sum^{N_{cap}}_{n=1}q_n.
    \label{conservation_mass}
\end{eqnarray}
That is, the total flow entering capillary bed from arteriolar tree equals to the total flow drained into venules.
This consistency condition could also be derived by integrating the solution for $p_{mean}$ over the boundary:
\begin{eqnarray}
    \overline{p_{mean}}=-\frac{1}{S}\frac{1}{\kappa_{cap}+\kappa_t}\sum^{N_{cap}}_{n=1}q_n\int_{\partial \Omega_t}f_{mean}(r,\theta;r_n,\theta_n)ds+\overline{p_{mean}},
\end{eqnarray}
which implies that
\begin{eqnarray}
    0=-\frac{1}{S}\frac{1}{\kappa_{cap}+\kappa_t}\sum^{N_{cap}}_{n=1}q_n\int_{\partial \Omega_t}f_{mean}(r,\theta;r_n,\theta_n)ds.
    \label{conservation_mass_bounary_average}
\end{eqnarray}
This demonstrates that the total flow is balanced, with each $q_n$ weighted by its respective boundary integral.
In other words the boundary integral quantifies how any local flow source disturbs the averaged pressure.
On the boundary $\partial\Omega_t$, where $r=R_t$, the function $f_{mean}$ simplifies to
\begin{eqnarray}
    f_{mean}(R_t,\theta;r_n,\theta_n)=\frac{1}{2\pi}\log(R^2_t+r^2_n-2R_tr_n\cos(\theta-\theta_n)).
\end{eqnarray}
The boundary integral can be computed using the identity that \cite[4.224.9]{gradshtein_table_2015}
\begin{eqnarray}
    \int^{\pi}_0\log(a+b\cos \theta)d\theta=\pi\log\frac{a+\sqrt{a^2-b^2}}{2}, \quad a\geq \vert b\vert>0.
    \label{logarithmic_inetgral}
\end{eqnarray}
Interestingly, evaluation of the boundary integral shows it is independent of the source location:
\begin{eqnarray}
    \int_{\partial \Omega_t}f_{mean}(r,\theta;r_n,\theta_n)ds=2R_t\log R_t.
\end{eqnarray}
Thus the equation$~$(\ref{conservation_mass_bounary_average}) can be simplified to equation$~$(\ref{conservation_mass}).
The identical value of boundary integral for each source demonstrates that each connection to the larger vasculature, whether a feeding arteriole or a draining venule, contributes equally to the balance, which is consistent with the physiological situation that capillary plexus is highly interconnected.

The problem of the Green's function $f_{exch}$ for $p_{exch}$ is given by
\begin{eqnarray}
    \left\{
    \begin{aligned}
        &\nabla^2f_{exch}({\bm x},{\bm x}_n)-\lambda^2 f_{exch}({\bm x},{\bm x}_n)=\delta({\bm x}-{\bm x}_n)\quad &\text{in $\Omega_t$}, \\
        &\nabla f_{exch}({\bm x},{\bm x}_n)\cdot {\bm n}=0\quad &\text{on $\partial\Omega_t$},
        \label{p_exch_unit_source}
    \end{aligned}\right.
\end{eqnarray}
whose solution enables to express $p_{exch}$ as
\begin{eqnarray}
    p_{exch}({\bm x})=\frac{1}{\kappa_{cap}+\kappa_t}\sum^{N_{cap}}_{n=1}q_nf_{exch}({\bm x},{\bm x}_n).
    \label{adjoint_problem_p_exch}
\end{eqnarray}
In polar coordinates, the equation becomes
\begin{eqnarray}
    \frac{1}{r}\frac{\partial}{\partial r}\left(r\frac{\partial f_{exch}}{\partial r}\right)+\frac{1}{r^2}\frac{\partial^2 f_{exch}}{\partial \theta^2}-\lambda^2f_{exch}=\frac{\delta(r-r_n)\delta(\theta-\theta_n)}{r},
    \label{radial_angle}
\end{eqnarray}
and substituting the trial solution $f_{exch}(r,\theta;r_n,\theta_n)=R(r)\Theta(\theta)$ leads to
\begin{eqnarray}
    \frac{r}{R(r)}\frac{\partial}{\partial r}\left(r\frac{\partial R(r)}{\partial r}\right)+\frac{1}{\Theta(\theta)}\frac{\partial^2 \Theta(\theta)}{\partial \theta^2}-\lambda^2r^2=0
\end{eqnarray}
as $(r,\theta)\neq (r_n,\theta_n)$, which yields
\begin{eqnarray}
    \frac{d^2\Theta}{d\theta^2}+m^2\Theta=0,
\end{eqnarray}

where periodicity in $\theta$ (with period $2\pi$) requires that $m\in \mathbb{Z}$ and its solution is composed of $\exp(im\theta)$ and $\exp(-im\theta)$.
Together with the Fourier expansion of $\delta(\theta-\theta_n)$ that
\begin{eqnarray}
    \delta(\theta-\theta_n)=\frac{1}{2\pi}\sum^\infty_{m=-\infty}\exp[im(\theta-\theta_n)],
\end{eqnarray}
we thus assume the solution is
\begin{eqnarray}
    f_{exch}(r,\theta;r_n,\theta_n)=\frac{1}{2\pi}\sum^{\infty}_{m=-\infty}R_{m,n}(r)\exp[im(\theta-\theta_n)],
    \label{f_exch_solution}
\end{eqnarray}
which yields following eigenvalue problem
\begin{eqnarray}
    \frac{1}{r}\frac{d}{dr}\left(r\frac{dR_{m,n}}{dr}\right)-\left(\frac{m^2}{r^2}+\lambda^2\right)R_{m,n}=\frac{\delta(r-r_n)}{r}.
    \label{radial_func}
\end{eqnarray}
The equation$~$(\ref{radial_func}) is a modified Bessel equation and its general solution, constrained to be finite at $r=0$, is given by
\begin{eqnarray}
    R_{m,n}(r)=\left\{
    \begin{aligned}
        &A_{m,n}I_m(\lambda r),\quad &r\leq r_n,\\
        &B_{m,n}I_m(\lambda r)+C_{m,n}K_m(\lambda r),\quad &r>r_n,
    \end{aligned}\right.
    \label{radial_component_func}
\end{eqnarray}
where $I_m(\lambda r)$ and $K_m(\lambda r)$ are the modified Bessel functions of order $m$ and $A_{m,n}$, $B_{m,n}$ as well as $C_{m,n}$ are the coefficients left to determine, and a detailed computation for them is provided in Appendix$~$\ref{SolutionExchangePressure}.
The radial component in equation$~$(\ref{f_exch_solution}) is given by
\begin{eqnarray}
    R_{m,n}(r)=I_m(\lambda r^n_<)\left[\frac{K^\prime_m(\lambda R_t)}{I^\prime_m(\lambda R_t)}I_m(\lambda r^n_>)-K_m(\lambda r^n_>)\right],
    \label{radial_component}
\end{eqnarray}
with $r^n_<=\min(r,r_n)$ and $r^n_>=\max(r,r_n)$.
Plugging the expression of $f_{exch}$ with equation$~$(\ref{radial_component}) into equation$~$(\ref{adjoint_problem_p_exch}) yields
\begin{eqnarray}
    p_{exch}(r,\theta)=\frac{1}{2\pi(\kappa_{cap}+\kappa_t)}\sum^{N_{cap}}_{n=1}\sum^{\infty}_{m=-\infty}q_nR_{m,n}(r)\exp[im(\theta-\theta_n)].
    \label{solution_p_exch}
\end{eqnarray}

With Green's functions for mean and exchange pressures,
we denote the modified Green's functions for capillary and tissue pressures as
\begin{eqnarray}
    G_{cap}(r,\theta;r_n,\theta_n)=f_{mean}(r,\theta;r_n,\theta_n)+\frac{\kappa_t}{\kappa_{cap}}f_{exch}(r,\theta;r_n,\theta_n), 
    \label{Greens_func_cap} \\
    G_t(r,\theta;r_n,\theta_n)=f_{mean}(r,\theta;r_n,\theta_n)-f_{exch}(r,\theta;r_n,\theta_n),
    \label{Greens_func_t}
\end{eqnarray}
where the functions $f_{mean}$ and $f_{exch}$ are given by equations$~$(\ref{f_mean_solution}) and (\ref{f_exch_solution}), respectively, and the pressures can then be expressed as
\begin{eqnarray}
    p_{cap}(r,\theta)=-\frac{1}{\kappa_{cap}+\kappa_t}\sum^{N_{cap}}_{n=1}q_nG_{cap}(r,\theta;r_n,\theta_n)+\overline{p_{mean}}+\alpha_r\Delta\pi, \label{capillary_pressure} \\
    p_t(r,\theta)=-\frac{1}{\kappa_{cap}+\kappa_t}\sum^{N_{cap}}_{n=1}q_nG_t(r,\theta;r_n,\theta_n)+\overline{p_{mean}}.
    \label{tissue_pressure}
\end{eqnarray}
\begin{figure}
	\centering
	\includegraphics[width=0.86\textwidth]{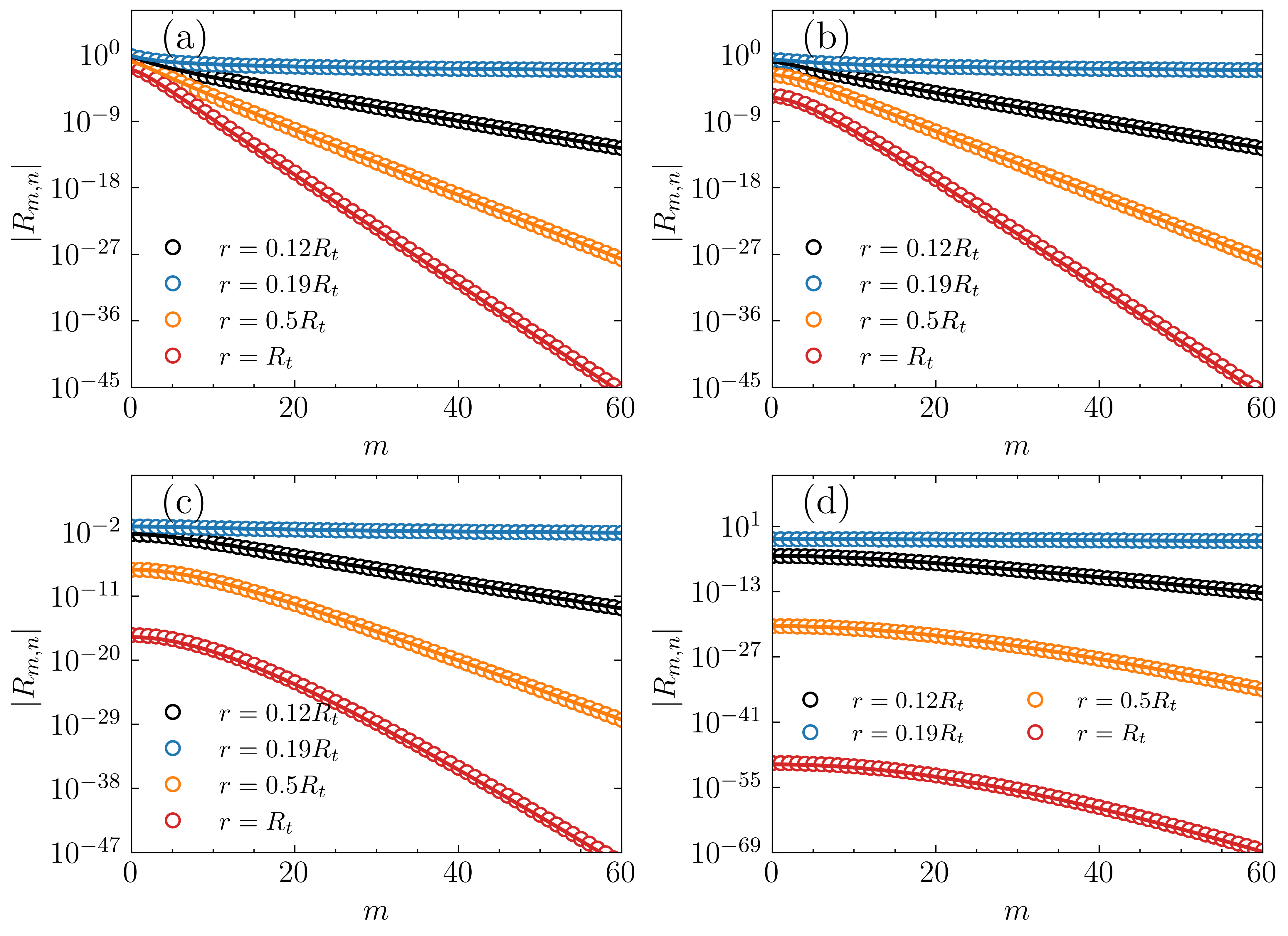}
	\caption{Dependence of radial component $R_{m,n}$ on the order $m$ for a source located at $r_n=0.19R_t$. Circles are the exact values of radial component computed by equation$~$(\ref{radial_component}) while lines denote the approximation$~$(\ref{radial_component_approximation}). Results are shown for four drainage rates: (a) $\alpha=2\times 10^{-5}~\rm{{\upmu m \cdot s/g}}$, (b) $\alpha=2\times 10^{-4}~\rm{{\upmu m \cdot s/g}}$, (c) $\alpha=2\times 10^{-3}~\rm{{\upmu m \cdot s/g}}$, and (d) $\alpha=2\times 10^{-2}~\rm{{\upmu m \cdot s/g}}$. Other parameter values are $k_{cap}=0.03~\rm{\upmu m}^2$ and $k_t=2\times 10^{-4}~\rm{\upmu m}^2$. Different colours correspond to various values of the radial coordinate of the source.}
	\label{R_m_n_large_m}
\end{figure}

\subsection{Convergence of the series solution for exchange pressure}
The solution for $p_{exch}$ is constructed by superimposing the contributions from all source points, where the kernel $f_{exch}$ at each source is expressed as an infinite series $~$(\ref{f_exch_solution}).
In order to analyse its convergence, $f_{exch}$ is decomposed into two components as
\begin{eqnarray}
	f_{exch}=s_1+s_2,
	\label{decomposition_f_exch}
\end{eqnarray}
where
\begin{eqnarray}
	s_1=\frac{1}{2\pi}\sum^{\infty}_{m=-\infty}I_m(\lambda r^n_<)\frac{K^\prime_m(\lambda R_t)}{I^\prime_m(\lambda R_t)}I_m(\lambda r^n_>)\exp[im(\theta-\theta_n)], \\
	s_2=-\frac{1}{2\pi}\sum^{\infty}_{m=-\infty}I_m(\lambda r^n_<)K_m(\lambda r^n_>)\exp[im(\theta-\theta_n)].
\end{eqnarray}
As discussed in Appendix$~$\ref{RadtioRadialComponent}, the terms in $s_1$ become negligible compared to those in $s_2$ for large $m$, which implies the following approximation for radial component:
\begin{eqnarray}
     R_{m,n}(r)\simeq -I_m(\lambda r^n_<)K_m(\lambda r^n_>),
     \label{radial_component_approximation}
\end{eqnarray}
and as a result $f_{exch}$ is effectively approximated by $s_2$.
The radial component depends on the parameter $\lambda=\sqrt{\alpha_{exch}(1/\kappa_{cap}+1/\kappa_t)}$, thus its value varies with $\alpha_{exch}$ and $1/\kappa_{cap}+1/\kappa_t$.
The accuracy of approximation$~$(\ref{radial_component_approximation}) is explored below for varying $\alpha_{exch}$ and fixed permeabilities.
\begin{figure}
	\centering
	\includegraphics[width=0.86\textwidth]{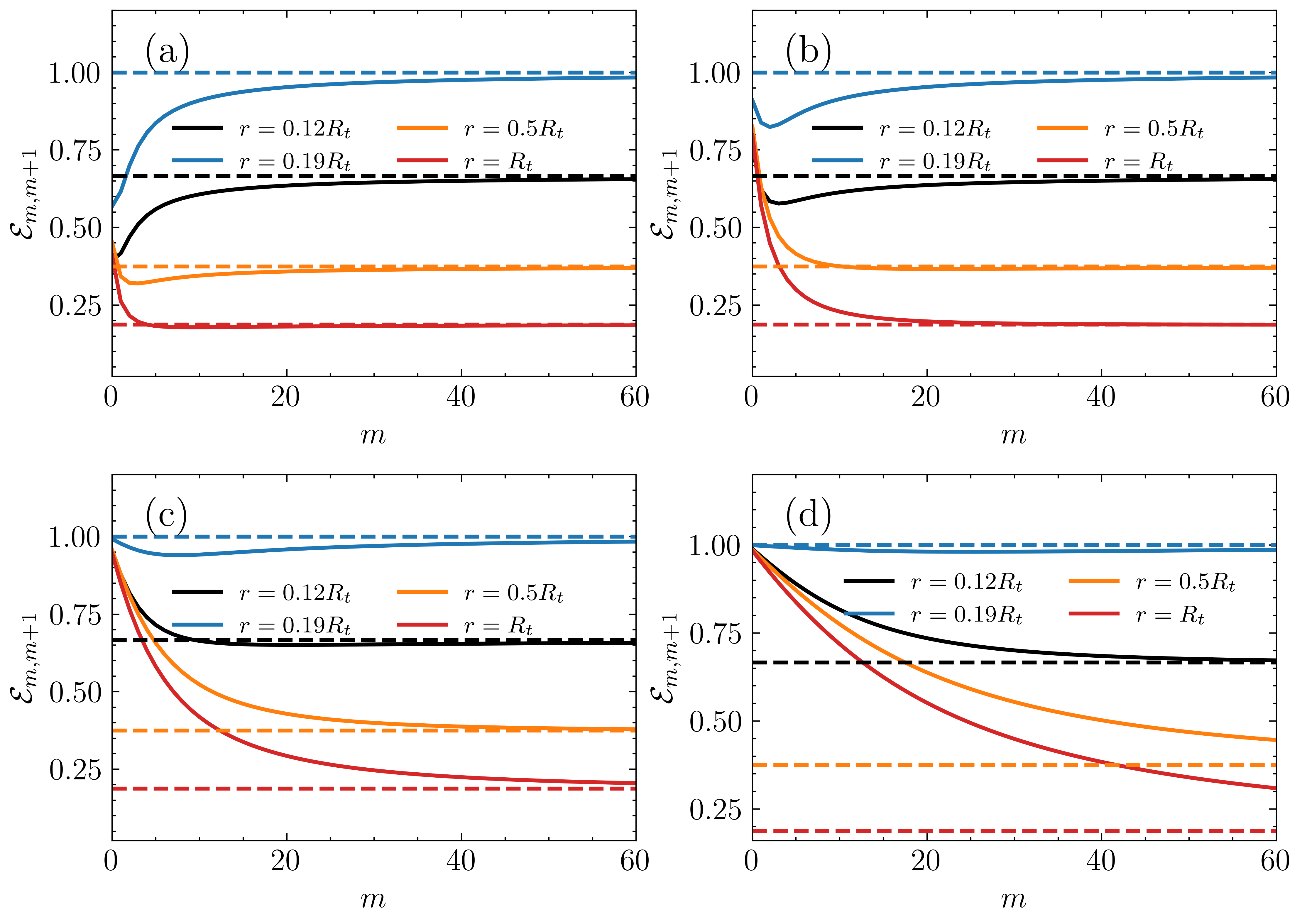}
	\caption{Dependence of relative ratio $\mathcal{E}_{m,m+1}$ on the order $m$ for a source located at $r_n=0.19R_t$. Solid lines denote the values of relative ratio while dashed lines are the upper bound. Results are shown for four drainage rates: (a) $\alpha=2\times 10^{-5}~\rm{{\upmu m \cdot s/g}}$, (b) $\alpha=2\times 10^{-4}~\rm{{\upmu m \cdot s/g}}$, (c) $\alpha=2\times 10^{-3}~\rm{{\upmu m \cdot s/g}}$, and (d) $\alpha=2\times 10^{-2}~\rm{{\upmu m \cdot s/g}}$. Other parameter values are $k_{cap}=0.03~\rm{\upmu m}^2$ and $k_t=2\times 10^{-4}~\rm{\upmu m}^2$. Different colours correspond to various values of the radial coordinate of the source.}
	\label{E_m_m_1_upper}
\end{figure}

Figure$~$\ref{R_m_n_large_m} displays the dependence of radial component on the order $m$, where circles denote the exact values from equation$~$(\ref{radial_component}) and lines represent the approximation$~$(\ref{radial_component_approximation}).
Subfigures (a), (b), (c) and (d) show the results for exchange rates of $\alpha=2\times 10^{-5}~{\rm{\upmu m}\cdot \rm{s}/\rm{g}}$, $\alpha=2\times 10^{-4}~{\rm{\upmu m}\cdot \rm{s}/\rm{g}}$, $\alpha=2\times 10^{-3}~{\rm{\upmu m}\cdot \rm{s}/\rm{g}}$, and $\alpha=2\times 10^{-2}~{\rm{\upmu m}\cdot \rm{s}/\rm{g}}$, respectively.
The approximation$~$(\ref{radial_component_approximation}) is nearly identical to the exact value across various values of radial variable, even yielding a good estimate at $r=R_t$.
These results demonstrate that the expression$~$(\ref{radial_component_approximation}) provides an excellent estimate for the radial component despite parameter variations, and thus the convergence of $f_{exch}$ is equivalent to that of series $s_2$.

\begin{figure}
	\centering
	\includegraphics[width=\textwidth]{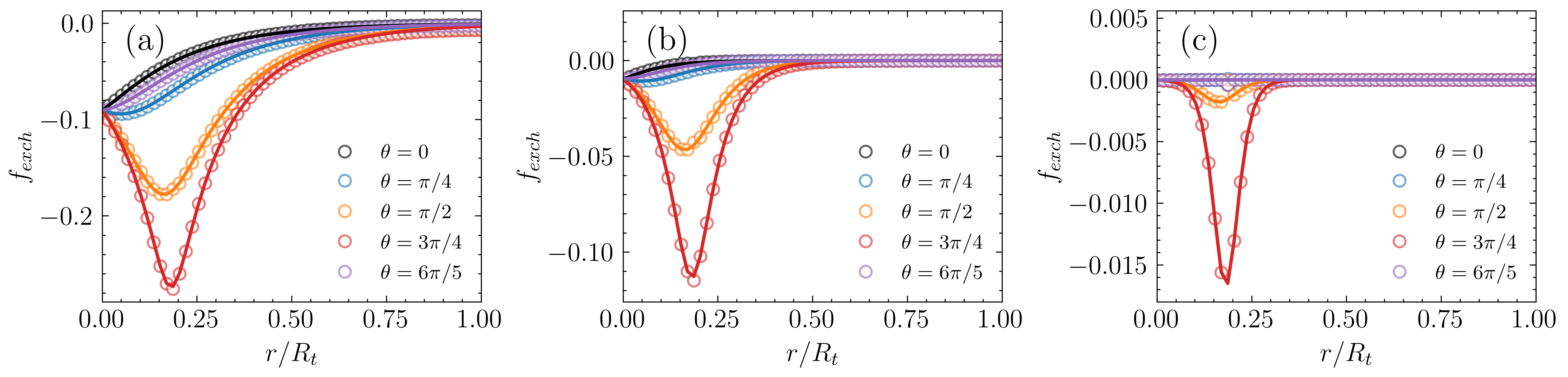}
	\caption{Behaviour of $f_2$ on $r/R_t$ with $r_n=0.19R_t$ and $\theta_n=2\pi/3$. Circles are results of the truncation$~$(\ref{f_exch_solution}) with order $M=26$, while lines are the values of approximation$~$(\ref{f_exch_approximation}). Different colours correspond to various values of $\theta$. Results are shown for four drainage rates: (a) $\alpha=2\times 10^{-5}~\rm{{\upmu m \cdot s/g}}$, (b) $\alpha=2\times 10^{-4}~\rm{{\upmu m \cdot s/g}}$, (c) $\alpha=2\times 10^{-3}~\rm{{\upmu m \cdot s/g}}$, and (d) $\alpha=2\times 10^{-2}~\rm{{\upmu m \cdot s/g}}$. Other parameter values are $k_{cap}=0.03~\rm{\upmu m}^2$ and $k_t=2\times 10^{-4}~\rm{\upmu m}^2$.}
	\label{f_exch_large_m}
\end{figure}
The convergence of $s_2$ can be established by the way of the following relative ratio
\begin{eqnarray}
    \mathcal{E}_{m,m+1}=\frac{K_{m+1}(\lambda r^n_>)}{K_m(\lambda r^n_>)}\frac{I_{m+1}(\lambda r^n_<)}{I_m(\lambda r^n_<)}.
\end{eqnarray}
Using the inequalities that \cite[Theorems 2 and 7]{ruiz-antolin_new_2016}
\begin{eqnarray}
    &&\frac{K_{m+1}(\lambda r)}{K_m(\lambda r)}<\frac{m+\frac{1}{2}+\sqrt{(m+\frac{1}{2})^2+(\lambda r)^2}}{\lambda r}, \quad m\geq -1/2\\
    &&\frac{I_{m+1}(\lambda r)}{I_m(\lambda r)}<\frac{\lambda r}{m+\frac{1}{2}+\sqrt{\left(m+\frac{1}{2}\right)^2+(\lambda r)^2}}, \quad m\geq -1/2,
\end{eqnarray}
we obtain
\begin{eqnarray}
    \mathcal{E}_{m,m+1}<\frac{m+\frac{1}{2}+\sqrt{(m+\frac{1}{2})^2+(\lambda r^n_>)^2}}{m+\frac{1}{2}+\sqrt{\left(m+\frac{1}{2}\right)^2+(\lambda r^n_<)^2}}\frac{r^n_<}{r^n_>},
\end{eqnarray}
thus for all $r$ and sufficiently large $m$ the ratio satisfies
\begin{eqnarray}
    \mathcal{E}_{m,m+1}<\frac{\min\{r,r_n\}}{\max\{r,r_n\}}\leq 1.
    \label{relative_ratio_upper}
\end{eqnarray}

Figure$~$\ref{E_m_m_1_upper} illustrates the dependence of relative ratio on the order $m$, where solid lines represent the values of relative ratio and dashed lines are the upper bound $\min\{r,r_n\}/\max\{r,r_n\}$, and different colours correspond to various values of radial coordinate.
Subfigures$~$(a) and (b) show the results for drainage rates of $\alpha=2\times 10^{-5}~{\rm{\upmu m}\cdot \rm{s}/\rm{g}}$ and $\alpha=2\times 10^{-4}~{\rm{\upmu m}\cdot \rm{s}/\rm{g}}$, respectively.
When $m$ increases, the relative ratio approaches the upper bound and remains strictly below it for large $m$.
Subfigure$~$(c) shows the result for drainage rates of $\alpha=2\times 10^{-3}~{\rm{\upmu m}\cdot \rm{s}/\rm{g}}$, the relative ratio converges to upper bound and remains strictly below it, except at $r=0.5R_t$ and $r=R_t$.
As drainage rate becomes larger, as shown in figure$~$\ref{E_m_m_1_upper}(d), the convergence of relative ratio toward the bound is slower, but the ratio is still less than $1$.

Since $s_2$ can be expressed as
\begin{eqnarray}
    s_2=-\frac{1}{2\pi}I_0(\lambda r^n_<)K_0(\lambda r^n_>)-\frac{1}{\pi}\sum^{\infty}_{m=1}I_m(\lambda r^n_<)K_m(\lambda r^n_>)\cos[m(\theta-\theta_n)],
\end{eqnarray}
using the equation$~$(\ref{relative_ratio_upper}) we obtain the following inequality:
\begin{eqnarray}
	\vert s_2\vert\leq \frac{1}{\pi}I_0(\lambda r^n_<)K_0(\lambda r^n_>)\left(\frac{1}{2}+\sum^\infty_{m=1}\left(\frac{\min\{r,r_n\}}{\max\{r,r_n\}}\right)^m\right),
\end{eqnarray}
and evaluating the series yields
\begin{eqnarray}
    \vert s_2\vert\leq\frac{1}{\pi}I_0(\lambda r^n_<)K_0(\lambda r^n_>)\frac{1+\min\{r,r_n\}/\max\{r,r_n\}}{2\left(1-\min\{r,r_n\}/\max\{r,r_n\}\right)},
\end{eqnarray}
which implies the convergence of $s_2$.
Thus $f_{exch}$ converges asymptotically, with the same rate as $s_2$.
The series $s_2$ can be written in a more compact expression using the following addition formula for modified Bessel functions:
\begin{eqnarray}
    K_0(\lambda \xi_n)=\sum^{\infty}_{m=-\infty}I_m(\lambda r^n_<)K_m(\lambda r^n_>)\exp[im(\theta-\theta_n)],
\end{eqnarray}
where $\xi_n=\sqrt{r^2+r^2_n-2rr_n\cos(\theta-\theta_n)}$ is the distance between field point and source point. 
This identity can be considered the counterpart of Graf addition formula for Bessel functions \cite{abramowitz_handbook_2013}, and is briefly proved in Appendix$~$\ref{ProofGrafAddition}.
The addition formula for modified Bessel functions implies that
\begin{eqnarray}
	f_{exch}(r,\theta;r_n,\theta_n)\simeq -\frac{1}{2\pi} K_0(\lambda \xi_n),
	\label{f_exch_approximation}
\end{eqnarray}
thus the series solution for $f_{exch}$ converges asymptotically.
Using the addition formula $f_{exch}$ can be written as
\begin{eqnarray}
    f_{exch}=-\frac{1}{2\pi}K_0(\lambda \xi_n)+\frac{1}{2\pi}\sum^{\infty}_{m=-\infty}I_m(\lambda r^n_<)\frac{K^\prime_m(\lambda R_t)}{I^\prime_m(\lambda R_t)}I_m(\lambda r^n_>)\exp[im(\theta-\theta_n)].
\end{eqnarray}

Figure$~$\ref{f_exch_large_m} illustrates the behaviour of $f_{exch}$ across $r/R_t$ for various values of parameters, where circles denote the truncation of $f_{exch}$ with truncation order $M=26$ and lines represent the free-space approximation$~$(\ref{f_exch_approximation}).
For three values of drainage rate, the free-space approximation agrees closely with the truncation across most of the domain.
However, for curves with $\theta$ close to $\theta_n$, such as the purple curve, the approximation deviates from the truncation.

These results, on the one hand, demonstrate the asymptotic convergence of $f_{exch}$. On the other hand, they also suggest how to model tissue with complicated, irregular shape.
The capillary-tissue coupled system can be decoupled into mean pressure $p_{mean}$ and exchange pressure $p_{exch}$.
The mean pressure, which might be sensitive to the boundary, can be solved numerically, while the  exchange pressure, which governs perfusion, can be effectively approximated using the free-space solution$~$(\ref{f_exch_approximation}).

\subsection{Domain-averaged pressures}
In addition to the spatial pressure fields, we compute the domain average of capillary and tissue pressures.
These global metrics could provide a compact summary of model behaviours and facilitate new experimental observations.

The domain-averaged pressures are given by
\begin{eqnarray}
    \langle p_{cap}\rangle=\frac{1}{A_t}\int_{\Omega_t} p_{cap}d{\bm x}=\frac{1}{\kappa_{cap}+\kappa_t}\sum^{N_{cap}}_{n=1}q_n\langle G_{cap}(r_n,\theta_n)\rangle+\overline{p_{mean}}+\alpha_r\Delta\pi, \\
    \langle p_t\rangle=\frac{1}{A_t}\int_{\Omega_t} p_td{\bm x}=\frac{1}{\kappa_{cap}+\kappa_t}\sum^{N_{cap}}_{n=1}q_n\langle G_t(r_n,\theta_n)\rangle+\overline{p_{mean}},
\end{eqnarray}
where $A_t=\pi R^2_t$ is the area of tissue domain and
\begin{eqnarray}
    \langle G_{cap}(r_n,\theta_n)\rangle=\frac{1}{A_t}\int^{2\pi}_0\int^{R_t}_0 G_{cap}(r,\theta;r_n,\theta_n)rdrd\theta, \\
     \langle G_t(r_n,\theta_n)\rangle=\frac{1}{A_t}\int^{2\pi}_0\int^{R_t}_0 G_t(r,\theta;r_n,\theta_n)rdrd\theta,
\end{eqnarray}
which vary with source point $(r_n,\theta_n)$.
The domain averages $ \langle G_{cap}(r_n,\theta_n)$ and $\langle G_t(r_n,\theta_n)$ can be expressed as
\begin{eqnarray}
    \langle G_{cap}(r_n,\theta_n)\rangle=\langle f_{mean}(r_n,\theta_n)\rangle+\frac{\kappa_t}{\kappa_{cap}}\langle f_{exch}(r_n,\theta_n)\rangle, \\
    \langle G_t(r_n,\theta_n)\rangle=\langle f_{mean}(r_n,\theta_n)\rangle-\langle f_{exch}(r_n,\theta_n)\rangle,
\end{eqnarray}
where domain averages $\langle f_{mean}(r_n,\theta_n)\rangle$ and $\langle f_{exch}(r_n,\theta_n)\rangle$ vary with $(r_n,\theta_n)$ as well.
From the problem$~$(\ref{p_exch_unit_source}) the domain average $\langle f_{exch}(r_n,\theta_n)\rangle$ is computed as follows
\begin{eqnarray}
    \langle f_{exch}(r_n,\theta_n)\rangle&=&\frac{1}{\lambda^2A_t}\int_{\Omega_t}\nabla^2f_{exch}d{\bm x}-\frac{1}{\lambda^2A_t} \nonumber\\
    &=&\frac{1}{\lambda^2A_t}\int_{\partial \Omega_t}\nabla f_{exch}\cdot {\bm n}ds-\frac{1}{\lambda^2A_t} \nonumber\\
    &=&-\frac{1}{\lambda^2A_t}.
\end{eqnarray}
The domain average $\langle f_{mean}(r_n,\theta_n)\rangle$ is given by
\begin{eqnarray}
    \langle f_{mean}(r_n,\theta_n)\rangle=a_1+a_2,
\end{eqnarray}
where
\begin{eqnarray}
    a_1=\frac{1}{4\pi A_t}\int^{2\pi}_0\int^{R_t}_0\log(r^2_n+r^2-2r_nr\cos(\theta_n-\theta))rdrd\theta, \\
    a_2=\frac{1}{4\pi A_t}\int^{2\pi}_0\int^{R_t}_0\log\left(r^2_n+\frac{R^4_t}{r^2}-2r_n\frac{R^2_t}{r}\cos(\theta_n-\theta)\right)rdrd\theta.
\end{eqnarray}
Evaluating $a_1$ and $a_2$ using the equation$~$(\ref{logarithmic_inetgral}) yields
\begin{eqnarray}
    a_1=\frac{1}{2\pi}\log R_t-\frac{1}{4\pi}+\frac{r^2_n}{4\pi R^2_t}, \\
    a_2=\frac{1}{2\pi}\log R_t+\frac{1}{4\pi}.
\end{eqnarray}
Thus,
\begin{eqnarray}
     \langle f_{mean}(r_n,\theta_n)\rangle=\frac{1}{\pi}\log R_t+\frac{r^2_n}{4\pi R^2_t},
\end{eqnarray}
and
\begin{eqnarray}
    \langle G_{cap}(r_n,\theta_n)\rangle=\frac{1}{\pi}\log R_t+\frac{r^2_n}{4\pi R^2_t}-\frac{\kappa_t}{\kappa_{cap}}\frac{1}{\lambda^2\pi R^2_t}, \\
    \langle G_t(r_n,\theta_n)\rangle=\frac{1}{\pi}\log R_t+\frac{r^2_n}{4\pi R^2_t}+\frac{1}{\lambda^2\pi R^2_t},
\end{eqnarray}
then domain-averaged pressures are given by
\begin{eqnarray}
    \langle p_{cap}\rangle=\frac{1}{\pi(\kappa_{cap}+\kappa_t)}\sum^{N_{cap}}_{n=1}q_n\left(\log R_t+\frac{r^2_n}{4R^2_t}-\frac{\kappa_t}{\kappa_{cap}}\frac{1}{\lambda^2 R^2_t}\right)+\overline{p_{mean}}+\alpha_r\Delta\pi,\label{p_cap_domain_average},
\end{eqnarray}
and
\begin{eqnarray}
	\langle p_t\rangle=\frac{1}{\pi(\kappa_{cap}+\kappa_t)}\sum^{N_{cap}}_{n=1}q_n\left(\log R_t+\frac{r^2_n}{4R^2_t}+\frac{1}{\lambda^2R^2_t}\right)+\overline{p_{mean}}.\label{p_t_domain_average}
\end{eqnarray}
These expressions reveal that vessels contribute disproportionately to global retinal fluid balance.
Moreover, the permeability ratio $\kappa_t/\kappa_{cap}$ plays a crucial role in $\langle p_{cap}\rangle$ but does not affect $\langle p_t\rangle$.

\section{Numerical methods}
\label{NumericaMethods}
\subsection{Finite volume method for 1D model}
\begin{figure}
	\centering
	\includegraphics[width=\textwidth]{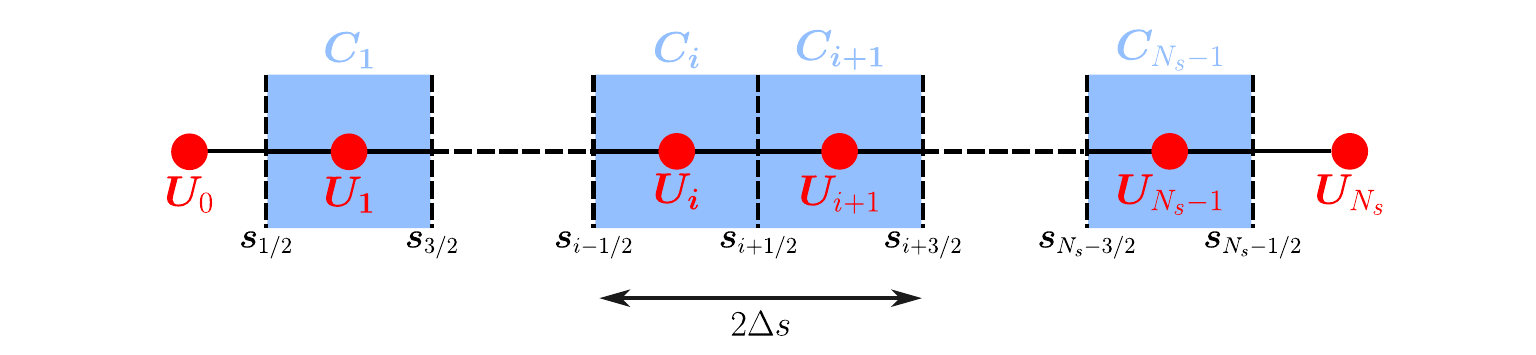}
	\caption{Spatial discretization of vessel.}
	\label{fvm}
\end{figure}

We solve the 1D blood flow model using the finite volume method (FVM) \cite{kolachalama_predictive_2007,olufsen_numerical_2000}.
The governing equations are expressed in conservative form,
\begin{eqnarray}
    \frac{\partial {\bm U}}{\partial t}+\frac{\partial {\bm F}}{\partial s}={\bm S},
    \label{conservative_form}
\end{eqnarray}
where state, flux and source vectors are given by
\begin{eqnarray}
    &&{\bm U}=(A,q)^T, \\
    &&{\bm F}=\left(q,\frac{4}{3}\frac{q^2}{A}+\frac{K}{\rho}\left(\frac{\beta_1}{\beta_1+1}\frac{A^{\beta_1+1}}{A^{\beta_1}_0}-\frac{\beta_2}{\beta_2+1}\frac{A^{\beta_2+1}}{A^{\beta_2}_0}\right)\right)^T, \\
    &&{\bm S}=\left(0,C_f\right)^T,
\end{eqnarray}
respectively.
The vessel domain $0\leq s\leq l_v$ is discretized into $N_s+1$ spatial points at $s_i=i\Delta s$ with $i=0,\dots,N_s$ and $\Delta s=l_v/N_s$.
The cell centres are located at $s_i$ with $1\leq i\leq N_s-1$ and the cells are $C_i=[s_{i-1/2},s_{i+1/2}]=[(i-1/2)\Delta s,(i+1/2)\Delta s]$, and the boundaries are at $s_0=0$ and $s_{N_s}=l_v$, as shown in figure$~$\ref{fvm}.
The time is discretized into steps $t_n=n\Delta t$ with $0\leq n\leq N_t$.The FVM is implemented on $C_i$ over $[t_n,t_{n+1}]$ while boundary conditions are enforced at $s_0$ and $s_{N_s}$.

We next describe the FVM for interior points.
Integrating equation$~$(\ref{conservative_form}) over the control volume $C_i\times[t_n,t_{n+1}]$ yields
\begin{eqnarray}
    \int^{t_{n+1}}_{t_n}\int_{C_i}{\bm S}(s,t)dsdt=&&\int_{C_i}\left[{\bm U}(s,t_{n+1})-{\bm U}(s,t_n)\right]ds\nonumber\\
    &&+\int^{t_{n+1}}_{t_n}\left[{\bm F}(s_{i+1/2},t)-{\bm F}(s_{i-1/2},t)\right]dt.
    \label{FVM_integral}
\end{eqnarray}
The integrals can be approximated as
\begin{eqnarray}
    \int^{s_{i+1/2}}_{s_{i-1/2}}{\bm U}(s,t_{n+1})ds&\approx&\Delta s{\bm U}^{n+1}_i,\\
    \int^{t_{n+1}}_{t_n}{\bm F}(s_{i+1/2},t)dt&\approx&\Delta t{\bm F}^{n+1/2}_{i+1/2},\\
    \int^{t_{n+1}}_{t_n}\int^{s_{i+1/2}}_{s_{i-1/2}}{\bm S}(s,t)dsdt&\approx&\frac{\Delta s\Delta t}{2}\left({\bm S}^{n+1/2}_{i+1/2}+{\bm S}^{n+1/2}_{i-1/2}\right),
\end{eqnarray}
where ${\bm U}^{n+1}_i={\bm U}(s_i,t_{n+1})$, and similar notations are used for flux and source vectors.
Thus FVM equation becomes
\begin{eqnarray}
    {\bm U}^{n+1}_{i}={\bm U}^n_i-\frac{\Delta t}{\Delta s}\left({\bm F}^{n+1/2}_{i+1/2}-{\bm F}^{n+1/2}_{i-1/2}\right)+\frac{\Delta t}{2}\left({\bm S}^{n+1/2}_{i+1/2}+{\bm S}^{n+1/2}_{i-1/2}\right).
    \label{fvm_flow_rate_area}
\end{eqnarray}
The unknown interface variables ${\bm F}^{n+1/2}_{i+1/2}$, ${\bm F}^{n+1/2}_{i-1/2}$, ${\bm S}^{n+1/2}_{i+1/2}$ and ${\bm S}^{n+1/2}_{i-1/2}$ depend on ${\bm U}^{n+1/2}_{m}$ with $m=i\pm1/2$, which can be determined by the Lax-Wendroff scheme as \cite{kolachalama_predictive_2007}
\begin{eqnarray}
    {\bm U}^{n+1/2}_{m}=\frac{{\bm U}^n_{m+1/2}+{\bm U}^n_{m-1/2}}{2}+\frac{\Delta t}{2}\left(-\frac{{\bm F}^n_{m+1/2}-{\bm F}^n_{m-1/2}}{\Delta s}+\frac{{\bm S}^n_{m+1/2}+{\bm S}^n_{m-1/2}}{2}\right).
    \label{lax_wendroff}
\end{eqnarray}
\subsection{Implementations for boundary and multiscale coupling conditions}
\subsubsection{Boundary conditions for vascular trees}
At the CRA inlet, the area $A^{n+1}_{CRA,0}$ is determined by solving the equation
\begin{eqnarray}
   p_{in,CRA}(t_{n+1})-p_{ext}=K\left[\left(\frac{A^{n+1}_{CRA,0}}{A_0}\right)^{\beta_1}-\left(\frac{A^{n+1}_{CRA,0}}{A_0}\right)^{\beta_2}\right],
   \label{inlet_boundary_area_CRA}
\end{eqnarray}
where $p_{in,CRA}(t_{n+1})$ is the prescribed CRA inlet pressure at time $t_{n+1}$.
The corresponding flow rate is computed by discretizing the mass-conservation equation \cite{he_one-dimensional_2004}
\begin{eqnarray}
    \frac{(A^{n+1}_{CRA,0}+A^{n+1}_{CRA,1})-(A^n_{CRA,0}+A^n_{CRA,1})}{2\Delta t}+\frac{q^{n+1}_{CRA,1}-q^{n+1}_{CRA,0}}{\Delta s_{CRA}}=0,
\end{eqnarray}
which yields the following equation
\begin{eqnarray}
    q^{n+1}_{CRA,0}=q^{n+1}_{CRA,1}+\frac{\Delta s_{CRA}}{2\Delta t}\left[(A^{n+1}_{CRA,0}+A^{n+1}_{CRA,1})-(A^n_{CRA,0}+A^n_{CRA,1})\right].
    \label{inlet_boundary_flow_rate_CRA_inlet}
\end{eqnarray}
At the CRV outlet, the area $A^{n+1}_{CRV,0}$ is obtained by solving the equation
\begin{eqnarray}
    p_{out,CRV}-p_{ext}=K\left[\left(\frac{A_{CRV,0}}{A_0}\right)^{\beta_1}-\left(\frac{A_{CRV,0}}{A_0}\right)^{\beta_2}\right],
    \label{outlet_boundary_area_CRV}
\end{eqnarray}
where $p_{out,CRV}$ is the CRV outlet pressure.
The flow rate at CRV outlet is determined by
\begin{eqnarray}
    q^{n+1}_{CRV,0}=q^{n+1}_{CRV,1}.
    \label{outlet_boundary_flow_rate_CRV}
\end{eqnarray}

For the bifurcation boundary, the mass-conservation equations are discretized, which yields following system at $t=t_{n+1}$ \cite{he_one-dimensional_2004,koppl_dimension_2023},
\begin{eqnarray}
    {\bm Y}({\bm x}_b)=[Y_1({\bm x}_b),Y_2({\bm x}_b),\dots,Y_6({\bm x}_b)]={\bm 0},
    \label{bifurcation_boundary_area_flow_rate}
\end{eqnarray}
where
\begin{eqnarray}
    &Y_1=q^{n+1}_{M,N_s}-q^{n+1}_{d_1,0}-q^{n+1}_{d_2,0}, \\
    &Y_2=p^{n+1}_{M,N_s}(A^{n+1}_{M,N_s})-p^{n+1}_{d_1,0}(A^{n+1}_{d_1,0}), \\
    &Y_3=p^{n+1}_{M,N_s}(A^{n+1}_{M,N_s})-p^{n+1}_{d_2,0}(A^{n+1}_{d_2,0}), \\
    &Y_4=\frac{(A^{n+1}_{M,N_s}+A^{n+1}_{M,N_s-1})-(A^n_{M,N_s}+A^n_{M,N_s-1})}{2\Delta t}+\frac{q^{n+1}_{M,N_s}-q^{n+1}_{M,N_s-1}}{\Delta s_M}, \\
    &Y_5=\frac{(A^{n+1}_{d_1,0}+A^{n+1}_{d_1,1})-(A^n_{d_1,0}+A^n_{d_1,1})}{2\Delta t}+\frac{q^{n+1}_{d_1,1}-q^{n+1}_{d_1,0}}{\Delta s_{d_1}},\\
    &Y_6=\frac{(A^{n+1}_{d_2,0}+A^{n+1}_{d_2,1})-(A^n_{d_2,0}+A^n_{d_2,1})}{2\Delta t}+\frac{q^{n+1}_{d_2,1}-q^{n+1}_{d_2,0}}{\Delta s_{d_2}},
\end{eqnarray}
and the pressures $p^{n+1}_{M,N_s}(A^{n+1}_{M,N_s})$, $p^{n+1}_{d_1,0}(A^{n+1}_{d_1,0})$ and $p^{n+1}_{d_2,0}(A^{n+1}_{d_2,0})$ depend on the corresponding areas through elastic model.
The unknown variables are
\begin{eqnarray}
    {\bm x}_b=[A^{n+1}_{M,N_s},q^{n+1}_{M,N_s},A^{n+1}_{d_1,0},q^{n+1}_{d_1,0},A^{n+1}_{d_2,0},q^{n+1}_{d_2,0}]^T,
\end{eqnarray}
while the interior variables of each vessel, such as $A^{n+1}_{M,N_s-1}$ and $q^{n+1}_{M,N_s-1}$, are obtained by integration along each vessel.
This system is solved at each time step to determine the bifurcation boundary variables.

\subsubsection{Multiscale coupling condition}
\label{multiscale_coupling_condition}
We denote the vectors of pressure and flow rate of source points at $t=t_{n+1}$ as
\begin{eqnarray}
    &&{\bm p}^{n+1}_{cap}=[p^{n+1}_{cap}({\bm x}_1),\dots,p^{n+1}_{cap}({\bm x}_{N_{cap}})]^T, \\
    &&{\bm q}^{n+1}=[q^{n+1}({\bm x}_1),\dots,q^{n+1}({\bm x}_{N_{cap}})]^T,
\end{eqnarray}
respectively, where $p^{n+1}_{cap}({\bm x}_i)$ and $q^{n+1}({\bm x}_i)$ are the pressure and the flow rate at ${\bm x}={\bm x}_i$ respectively, and $N_{cap}$ is the number of superficial vessels connected to structured trees.
We assume Poiseuille flow in structured trees, thus $q^{n+1}({\bm x}_i)$ is also the flow rate in arteriolar or venular trees in superficial layer, which is connected to capillaries at ${\bm x}={\bm x}_i$.
We introduce the matrix ${\bm R}$, which is composed of the equivalent resistances of structured trees, given by
\begin{eqnarray}
    {\bm R}=
    \begin{bmatrix}
        R_{a,eq }({\bm x}_1) & & \\
        & \ddots & \\
        & & R_{a,eq}({\bm x}_i) \\
        & & & R_{v,eq}({\bm x}_{i+1}) \\
        & & & & \ddots \\
        & & & & & R_{v,eq}({\bm x}_{N_{cap}})
    \end{bmatrix},
\end{eqnarray}
where $R_{a,eq}({\bm x}_i)$ denotes the equivalent resistance of a structured tree connecting arteriolar tree to capillary continuum at ${\bm x}={\bm x}_i$ and $R_{v,eq}({\bm x}_k)$ denotes that of a structured tree connecting a venular tree to capillary continuum at ${\bm x}={\bm x}_k$.
The capillary pressures at source points are given by
\begin{eqnarray}
    {\bm p}^{n+1}_{cap}={\bm M}^{cap}{\bm q}^{n+1}+(\overline{p_{mean}}+\alpha_r\Delta\pi){\bm 1},
\end{eqnarray}
where ${\bm M}^{cap}$ is resistance matrix with element $M^{cap}_{iw}=-G_{cap}(r_i,\theta_i;r_w,\theta_w)/(\kappa_{cap}+\kappa_t)$, while the Poiseuille equation for structured trees yields the following equation
\begin{eqnarray}
    {\bm p}^{n+1}_{cap}={\bm p}^{n+1}-{\bm R}{\bm q}^{n+1},
    \label{Poiseuille_capillary_coupling_bc}
\end{eqnarray}
where the pressure vector of the terminal vessels in superficial vascular trees ${\bm p}^{n+1}$ is given by
\begin{eqnarray}
    {\bm p}^{n+1}=[p^{n+1}_{1,N_s},\dots,p^{n+1}_{N_{cap},N_s}]^T.
\end{eqnarray}
Thus the coupling condition is given by
\begin{eqnarray}
    \left\{
    \begin{aligned}
        &{\bm p}^{n+1}_{cap}={\bm M}^{cap}{\bm q}^{n+1}+(\overline{p_{mean}}+\alpha_r\Delta\pi){\bm 1}, \\
        &{\bm p}^{n+1}_{cap}={\bm p}^{n+1}-{\bm R}{\bm q}^{n+1}, \\
        &\overline{p_{mean}}=\frac{1}{S}\int_{\partial \Omega_t}p_{mean}ds, \\
        &p_{mean}(r,\theta)=-\frac{1}{\kappa_{cap}+\kappa_t}\sum^{N_{cap}}_{i=1}q_if_{mean}(r,\theta;r_i,\theta_i)+\overline{p_{mean}}, \\
        &\sum^{N_{cap}}_{n=1}q_n=0,
    \end{aligned}\right.
    \label{coupling_condition_capillary_poiseuille_boundary_average}
\end{eqnarray}
where the pressures ${\bm p}^{n+1}$ are governed by following equations in terminal vessels
\begin{eqnarray}
    \left\{
    \begin{aligned}
        &\pdv{A_k}{t}+\pdv{q_k}{s}=0,\quad k=1,\dots,N_{cap}, \\
        &p_k-p_{ext}=K\left[\left(\frac{A}{A_0}\right)^{\beta_1}-\left(\frac{A}{A_0}\right)^{\beta_2}\right],\quad k=1,\dots,N_{cap},\\
    \end{aligned}\right.
\end{eqnarray}
and $A_k$ and $q_k$ denote the area and flow rate of the terminal vessels in superficial vascular trees, respectively.

Solving the system$~$(\ref{coupling_condition_capillary_poiseuille_boundary_average}) directly is challenging due to the integral term, the conservation of mass and the nonlinear dependence of pressure on area.
To simplify it, we begin with combining the first two equations in system$~$(\ref{coupling_condition_capillary_poiseuille_boundary_average}), which yields the following pressure-flow relationship:
\begin{eqnarray}
    {\bm p}^{n+1}={\bm R}^{cap}{\bm q}^{n+1}+(\overline{p_{mean}}+\alpha_r\Delta\pi){\bm 1},
    \label{capillary_poiseuille_pressure_flow}
\end{eqnarray}
where ${\bm R}^{cap}={\bm M}^{cap}+{\bm R}$.
Solving for the flow rate vector yields
\begin{eqnarray}
    {\bm q}^{n+1}=({\bm R}+{\bm M}^{cap})^{-1}{\bm p}^{n+1}-(\overline{p_{mean}}+\alpha_r\Delta\pi)({\bm R}+{\bm M}^{cap})^{-1}{\bm 1},
\end{eqnarray}
and then imposing the conservation of mass leads to
\begin{eqnarray*}
    0={\bm 1}^T({\bm R}+{\bm M}^{cap})^{-1}{\bm p}^{n+1}-(\overline{p_{mean}}+\alpha_r\Delta\pi){\bm 1}^T({\bm R}+{\bm M}^{cap})^{-1}{\bm 1},
\end{eqnarray*}
which implies that
\begin{eqnarray}
    \overline{p_{mean}}=\frac{1}{s_{cap}}{\bm 1}^T({\bm R}+{\bm M}^{cap})^{-1}{\bm p}^{n+1}-\alpha_r\Delta\pi,
    \label{boundary_average}
\end{eqnarray}
with $s_{cap}={\bm 1}^T({\bm R}+{\bm M}^{cap})^{-1}{\bm 1}$.
Substituting the expression of $\overline{p_{mean}}$ into the equation$~$(\ref{capillary_poiseuille_pressure_flow}), we obtain
\begin{eqnarray}
    {\bm p}^{n+1}={\bm R}^{cap}{\bm q}^{n+1}+\frac{1}{s_{cap}}\left[{\bm 1}^T({\bm R}+{\bm M}^{cap})^{-1}{\bm p}^{n+1}\right]{\bm 1},
\end{eqnarray}
which consists of two unknown vectors ${\bm p}^{n+1}$ and ${\bm q}^{n+1}$.
The mass-conservation is discretized as that for CRA inlet and bifurcation boundary such that for each arteriole or venule,
\begin{eqnarray}
    q^{n+1}_{k,N_s}=q^{n+1}_{k,N_s-1}-\frac{\Delta s_k}{2\Delta t}\left[(A^{n+1}_{k,N_s-1}+A^{n+1}_{k,N_s})-(A^n_{k,N_s}+A^n_{k,N_s-1})\right],
\end{eqnarray}
where $k=1,\dots,N_{cap}$.
Given that flows in structured trees are assumed as Poiseuille flow, the flow rates are equal to those at boundaries of terminal vessels in superficial vascular trees, that is $q^{n+1}({\bm x}_k)=q^{n+1}_{k,N_s}$.
This enables to express ${\bm q}^{n+1}$ as
\begin{eqnarray}
    {\bm q}^{n+1}=-{\bm B}\left[({\bm A}^{n+1}_{N_s}+{\bm A}^{n+1}_{N_s-1})-({\bm A}^n_{N_s}+{\bm A}^n_{N_s-1})\right]+{\bm f}^{n+1},
    \label{capillary_flow_rate_vector}
\end{eqnarray}
where area vectors are given by
\begin{eqnarray}
    {\bm A}^{n+1}_{N_s}&=&[A^{n+1}_{1,N_s},\dots,A^{n+1}_{N_{cap},N_s}]^T, \\
    {\bm A}^{n+1}_{N_s-1}&=&[A^{n+1}_{1,N_s-1},\dots,A^{n+1}_{N_{cap},N_s-1}]^T, \\
    {\bm A}^n_{N_s}&=&[A^n_{1,N_s},\dots,A^n_{N_{cap},N_s}]^T, \\
    {\bm A}^n_{N_s-1}&=&[A^n_{1,N_s-1},\dots,A^n_{N_{cap},N_s-1}]^T,
\end{eqnarray}
the matrix ${\bm B}$ is given by
\begin{eqnarray}
    {\bm B}=
    \begin{bmatrix}
        \Delta s_1/(2\Delta t) & &  \\
        & \ddots &\\
        & & \Delta s_{N_{cap}}/(2\Delta t)
    \end{bmatrix},
\end{eqnarray}
and ${\bm f}^{n+1}$ is given by
\begin{eqnarray}
    {\bm f}^{n+1}=[q^{n+1}_{1,N_s-1},\dots,q^{n+1}_{N_{cap},N_s-1}]^T.
\end{eqnarray}
The numerical implementation for capillary-tissue coupled system is thus given by
\begin{eqnarray}
    \left\{
    \begin{aligned}
        &{\bm p}^{n+1}({\bm A}^{n+1}_{N_s})=-{\bm R}^{cap}{\bm B}({\bm A}^{n+1}_{N_s}+{\bm A}^{n+1}_{N_s-1}-{\bm A}^n_{N_s}-{\bm A}^n_{N_s-1})+{\bm R}^{cap}{\bm f}^{n+1}+\overline{p_{mean}}{\bm 1}, \\
        &p^{n+1}_{k,N_s}-p_{ext}=K\left[\left(\frac{A^{n+1}_{k,N_s}}{A_0}\right)^{\beta_1}-\left(\frac{A^{n+1}_{k,N_s}}{A_0}\right)^{\beta_2}\right],\quad k=1,\dots,N_{cap},
    \end{aligned}\right.
    \label{capillary_tissue_coupling_condition}
\end{eqnarray}
where the boundary average is given by equation$~$(\ref{boundary_average}).
It is interesting to note that the effect of $\alpha_r\Delta\pi$ is absorbed into the domain-average $\overline{p_{mean}}$ via the conservation of mass, and the flow rates at source points are thus independent of $\alpha_r\Delta\pi$.
In other words, a constant oncotic offset does not alter the pressure difference established by the fixed arteriolar and venular boundary conditions.

\subsubsection{Numerical treatments of infinite summation and the singularity}
Two main difficulties arise when implementing the Green's function for mean and exchange pressures: the infinite summation for exchange pressure, and the singularity when field point is a source point.
In what follows we show how to address these issues.

The finite order truncation of $s_1$ in $f_{exch}$ is utilized in computations.
The truncation of $s_1$ with order $M$ is given by
\begin{eqnarray}
    s_1\approx\frac{1}{2\pi}\sum^M_{m=-M}I_m(\lambda r^n_<)\frac{K^\prime_m(\lambda R_t)}{I^\prime_m(\lambda R_t)}I_m(\lambda r^n_>)\exp[im(\theta-\theta_n)],
\end{eqnarray}
since the radial component is symmetric in $m$ the truncation can be written as

\begin{eqnarray}
    s_1&\approx&\frac{1}{2\pi}I_0(\lambda r^n_<)\frac{K^\prime_0(\lambda R_t)}{I^\prime_0(\lambda R_t)}I_0(\lambda r^n_>) \nonumber\\
    &&+\frac{1}{\pi}\sum^M_{m=1}I_m(\lambda r^n_<)\frac{K^\prime_m(\lambda R_t)}{I^\prime_m(\lambda R_t)}I_m(\lambda r^n_>)\cos[m(\theta-\theta_n)].
\end{eqnarray}
The truncation could be accurate with only a few terms since the high order contributions to $s_1$ are negligible relative to $s_2$, as discussed in Appendix$~$\ref{RadtioRadialComponent}.

The singularity occurs at $r=r_n$ in both $f_{mean}(r,\theta;r_n,\theta_n)$ and $f_{exch}(r,\theta;r_n,\theta_n)$, which could be solved via local domain averaging.
The following averages are employed for logarithm function and modified Bessel function of second kind:
\begin{eqnarray}
    \langle \log\xi^2_n\rangle=\frac{1}{A_n} \int_{\Omega_n}\log\xi^2_nr dr d\theta, \\
    \langle K_0(\lambda\xi_n)\rangle=\frac{1}{A_n} \int_{\Omega_n}K_0(\lambda \xi_n)r dr d\theta,
\end{eqnarray}
where $\xi_n=\sqrt{r^2+r^2_n-2rr_n\cos(\theta-\theta_n)}$, $\Omega_n$ is a circular domain of radius $r_{v,n}$ centred at source point ${\bm x}_n$, and $A_n$ is its area.
Such averaging, on the one hand, provides a numerical treatment for the singularity, and on the other hand it is also physiologically reasonable, since the blood does not enter or exit the capillary bed at a point but through a vessel of finite radius $r_{v,n}$.
Moreover the pressure $p_{cap}$ is an averaged variable over some capillaries, and averaging over the vessel cross-section thus ensures the consistency in the coupling across scales.
The average for logarithmic function is given by
\begin{eqnarray}
    \langle \log\xi^2_n\rangle=\frac{2}{\pi r^2_{v,n}}\int^{2\pi}_0\int^{r_{v,n}}_0\xi\log \xi d\xi d\theta=2\log r_{v,n}-1.
\end{eqnarray}
The average for modified Bessel function of second kind can be written as
\begin{eqnarray}
    \langle K_0(\lambda\xi_n)\rangle=\frac{1}{\lambda^2 \pi r^2_{v,n}}\int^{2\pi}_0\int^{\lambda r_{v,n}}_0K_0(\xi)\xi d\xi d\theta,
\end{eqnarray}
and using the identity that $d(\xi K_1(\xi))/d\xi=-\xi K_0(\xi)$, it is given by
\begin{eqnarray}
    \langle K_0(\lambda\xi_n)\rangle=-\frac{2}{\lambda r_{v,n}}K_1(\lambda r_{v,n})+\frac{2}{\lambda^2 r^2_{v,n}}.
\end{eqnarray}

For $r=r_n$ the Green's function of source point $(r_n,\theta_n)$ for mean pressure is implemented as
\begin{eqnarray}
    f_{mean}(r_n,\theta_n;r_n,\theta_n)=\frac{1}{4\pi}\left[2\log r_{v,n}-1+\log\left(r^2_n+\frac{R^4_t}{r^2_n}-2R^2_t\cos(\theta_n-\theta)\right)\right],
\end{eqnarray}
while that for exchange pressure is
\begin{eqnarray}
    f_{exch}(r_n,\theta_n;r_n,\theta_n)&=&-\frac{2}{\lambda r_{v,n}}K_1(\lambda r_{v,n})+\frac{2}{\lambda^2 r^2_{v,n}}+\frac{1}{2\pi}I^2_0(\lambda r_n)\frac{K^\prime_0(\lambda R_t)}{I^\prime_0(\lambda R_t)} \nonumber\\
    &&+\frac{1}{\pi}\sum^M_{m=1}I^2_m(\lambda r_n)\frac{K^\prime_m(\lambda R_t)}{I^\prime_m(\lambda R_t)}\cos[m(\theta-\theta_n)].
\end{eqnarray}
\begin{table}   
    \centering
    \begin{tabular}{l l l}
         Parameter& Value& References\\
         \hline
         \textit{Viscosity model}\\
         $\eta_p$ & $1.2~\rm{cP}$& \cite{pries_resistance_1994} \\
         & & \cite{secomb_blood_2013} \\
         & & \cite{bappoo_viscosity_2017} \\
         \hline
         \textit{Elastic model}\\
         $E$& $10^6~\rm{g/cm/s^2}$& \cite{julien_one-dimensional_2023}\\
         & & \cite{ferrara_biomechanical_2021} \\
         $\rm{WLR}$& $0.16$ (arteriole)& \cite{arichika_effects_2015} \\
          & $0.12$ (venule) & \cite{ertop_evaluation_2023} \\
         IOP & $16~\rm{mmHg}$ & \cite{guidoboni_intraocular_2014} \\
         \hline
         \textit{L-system}\\
         Radius of CRA& $165\pm 15~\rm{\upmu m}$& \cite{dorner_calculation_2002}\\
         & & \cite{brown_physics-informed_2024} \\
         & & \cite{hernandez_linking_2024} \\
         Radius of CRV& $196\pm 15~\rm{\upmu m}$& \cite{riva_blood_1985}\\
         $\xi$& $36\pm 6$& \cite{brown_physics-informed_2024}\\
         $v_a$& $1$ ($d_v>50~\rm{\upmu m}$)& \cite{brown_physics-informed_2024}\\
         & $0.8$ ($d_v\leq50~\rm{\upmu m}$) & \\
         $\gamma$& $2.1$ ($d_v>50~\rm{\upmu m}$)& \cite{brown_physics-informed_2024} \\
         & $2$ ($d_v\leq50~\rm{\upmu m}$) & \\
         \hline
         \textit{Structured tree model}\\
         $\xi$& $23$& \cite{liu_computational_2009} \\
         & & \cite{julien_one-dimensional_2023} \\
         $v_a$& $0.8$& \cite{brown_physics-informed_2024} \\
         & & \cite{liu_computational_2009} \\
         $\gamma$& $2$& \cite{julien_one-dimensional_2023}\\
         \hline
         \textit{Capillary-tissue coupled system}\\
         $\mu_{cap}$ & computed by the viscosity model& \cite{pries_resistance_1994} \\
         $\mu_t$ & $0.7~\rm{cP}$& Estimated\\
         $R_t$& $1.6~\rm{cm}$& \cite{causin_blood_2016} \\
         $\alpha_r$ & $1$ & \cite{truskey_transport_2009} \\
         $\Delta\pi$ & $16~\rm{mmHg}$ & \cite{morissette_colloid_1977} \\
         \hline
         \textit{Boundary conditions at CRA and CRV}\\
         $p_{in,CRA}$ (steady) &$62~\rm{mmHg}$ & \cite{guidoboni_intraocular_2014} \\
          $p_{in,CRA}$ (pulsatile) &$45-77~\rm{mmHg}$ & \cite{rebhan_computational_2019} \\
          & & \cite{guidoboni_intraocular_2014} \\
         $p_{out,CRV}$ & $16~\rm{mmHg}$ & \cite{rebhan_computational_2019} \\
         & & \cite{causin_blood_2016} \\
         & & \cite{guidoboni_effect_2014} \\
    \end{tabular}
    \caption{Parameters for the hemodynamics in retinal microcirculation.}
    \label{parameters_retinal_circulation}
\end{table}

\subsection{Summary of numerical methods}
The numerical solution for the hemodynamics in retinal microcirculation at each time step is obtained as follows.
First, variables at interior points of each vessel are solved using equations$~$(\ref{fvm_flow_rate_area}) and (\ref{lax_wendroff}).
Second, the boundary conditions of each vessel are determined using equations$~$(\ref{inlet_boundary_area_CRA}), (\ref{inlet_boundary_flow_rate_CRA_inlet}), (\ref{outlet_boundary_area_CRV}), (\ref{outlet_boundary_flow_rate_CRV}) and (\ref{bifurcation_boundary_area_flow_rate}).
Third, the coupling condition$~$(\ref{capillary_tissue_coupling_condition}) is solved, from which the remaining undetermined boundary variables are computed.

\section{Validation of proposed multiscale model}
\label{ValidationProposedMultiscaleModel}
In the following section, we first establish the model's numerical robustness by exploring the dependence of total flow on truncation order.
To assess the accuracy of our model, we then validate the model's predictions against experimental data and those of established models.
Third, the ability of our model to predict pulsatile pressure is shown by imposing time-dependent CRA inlet pressure.
For all simulations the CRV outlet pressure is set as $p_{out,CRV}=16~\rm{mmHg}$.
For steady arteriolar pressure the CRA inlet pressure is set to a constant baseline value of $p_{in,CRA}=62~\rm{mmHg}$ which is the averaged pressure over one cardiac cycle approximately \cite{guidoboni_intraocular_2014}.
While values of other parameters in the 1D model are summarised in table$~$\ref{parameters_retinal_circulation}, parameters in capillary-tissue system are set as $k_{cap}=0.03~\rm{\upmu m}^2$, $k_t=0.0002~\rm{\upmu m}^2$ and $\alpha=2\times 10^{-8}~\rm{cm\cdot s/g}$.
Two different distributions for source points are employed in the computation of steady flow, which are shown in figure$~$\ref{CRA_inlet_pressure_large_capillary_tissue_pressure}(a) and (d), and they are denoted as source distribution 1 and source distribution 2, respectively.
In the computation for pulsatile inlet pressure only source distribution 1 is utilized.
\subsection{Steady flow}
\begin{figure}
    \centering
    \includegraphics[width=0.86\textwidth]{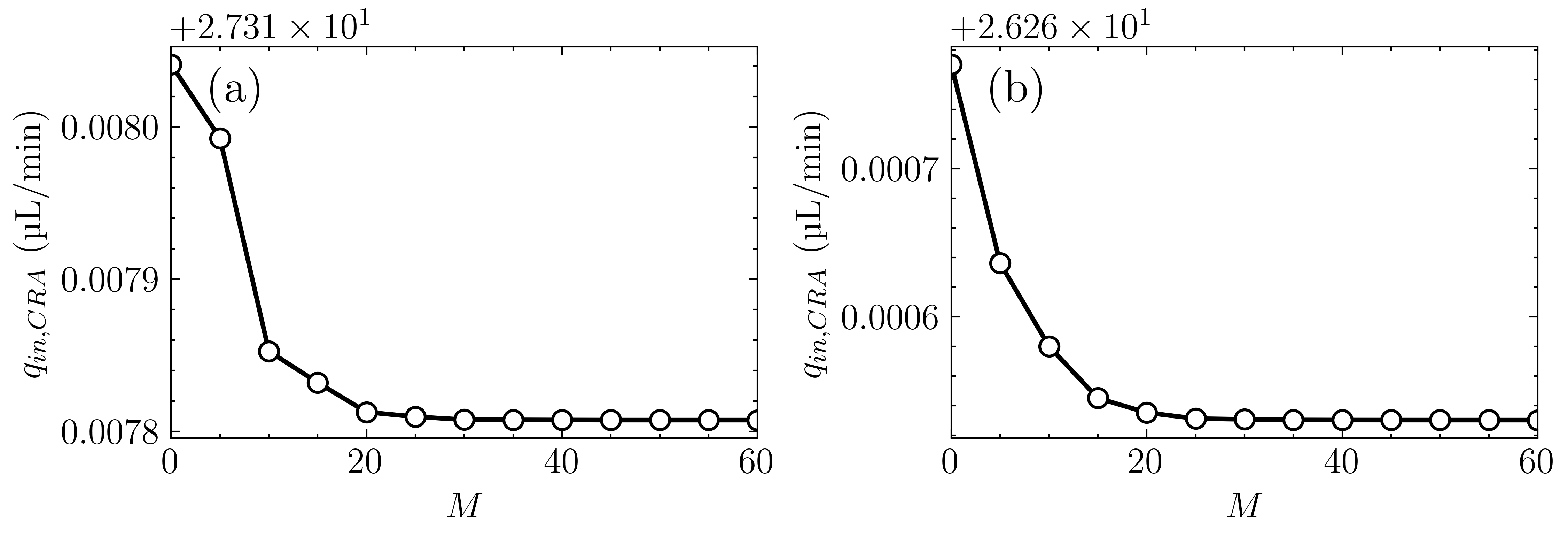}
    \caption{Dependence of total flow on truncation order $M$. (a) Result with source distribution 1. (b) Result with Source distribution 2.}
    \label{truncation_variation_flow}
\end{figure}

\begin{figure}
	\centering
	\includegraphics[width=\textwidth]{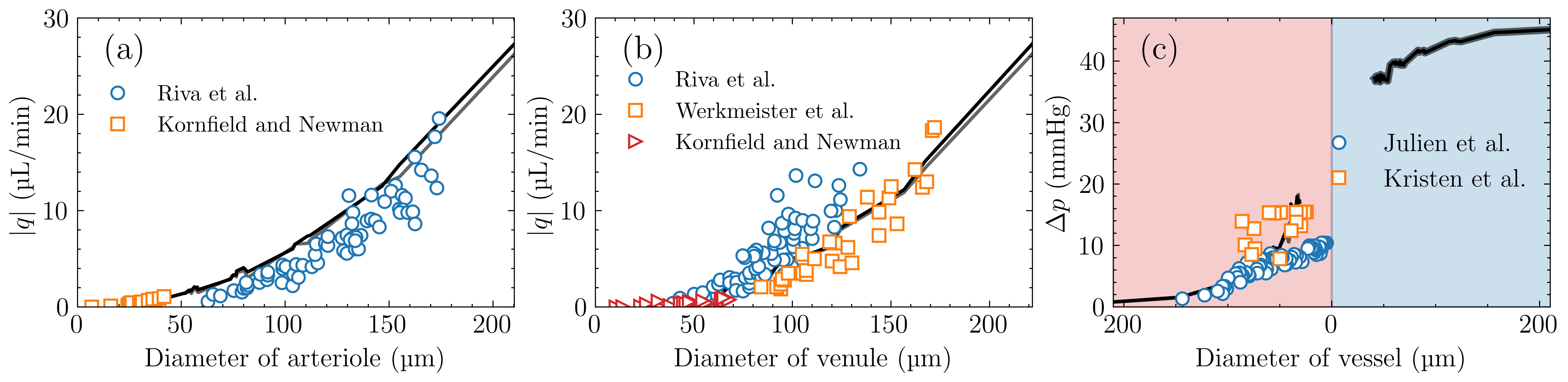}
	\caption{Distributions of flow rate and pressure difference in arterioles and venules. Lines denote the results of multiscale model and markers denote experimental data \cite{kornfield_measurement_2015,riva_blood_1985,werkmeister_measurement_2012} and predictions of other models \cite{julien_one-dimensional_2023,joldes_fundus_2016}. Black and grey lines denote the results with distribution 1 and distribution 2, respectively. (a) Flow rate distributions in arterioles. (b) Flow rate distributions in venules. (c) Pressure difference distributions in arterioles (red region) and venules (blue region). While arteriole diameter decreases from left to right in the red region (from CRA inlet to arterioles near capillary size), venule diameter increases from left to right in the blue region (from venules near capillary size to CRV outlet).}
	\label{CRA_inlet_pressure_variation_flow}
\end{figure}
We first explore the dependence of CRA inlet flow rate $q_{in,CRA}$ on the truncation order $M$, which is shown in figure$~$\ref{truncation_variation_flow}.
The value of $q_{in,CRA}$ exhibits convergence for both source distributions when $M>26$.
Besides, the variations are minimal, which are of approximately $2\times10^{-4}~\rm{\upmu L/min}$ across the entire range of $M$ for both distributions, which is about $7\times10^{-4}\%$ of total flow ($27.26~\rm{\upmu L/min}$ for source distribution 1 and $26.26~\rm{\upmu L/min}$ for source distribution 2).
Thus the model could be accurate with only a few terms in the truncation of series $s_1$.

\begin{figure}
	\centering
	\includegraphics[width=\textwidth]{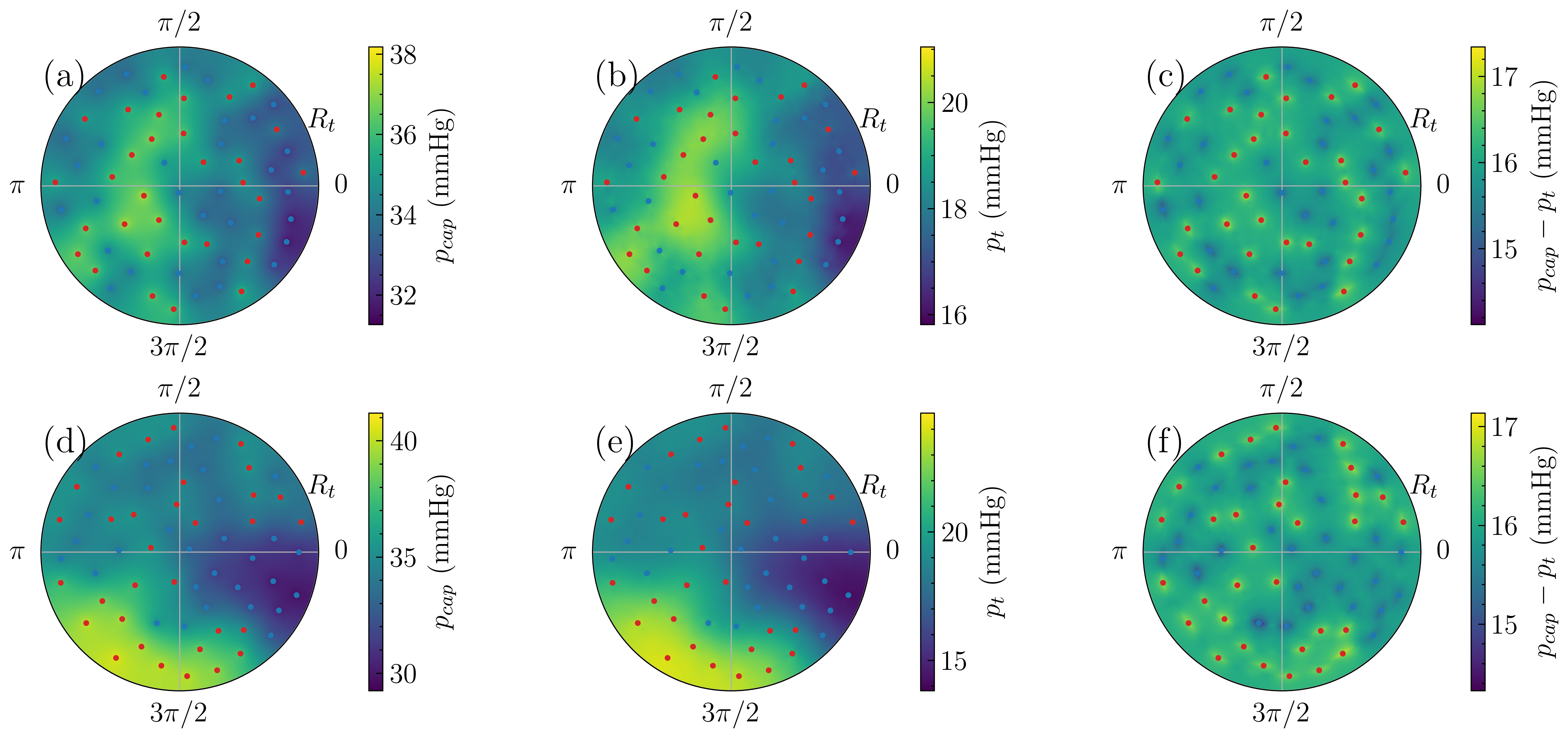}
	\caption{Capillary pressure, tissue pressure and exchange pressure. The first column correspond to the terminal source distribution while the second column correspond to the uniform source distribution. (a) and (d) Capillary pressures. (b) and (e) Tissue pressures. (c) and (f) Exchange pressures.}
	\label{CRA_inlet_pressure_large_capillary_tissue_pressure}
\end{figure}
Figure$~$\ref{CRA_inlet_pressure_variation_flow}(a) and (b) show the distributions of averaged flow rate along individual arterioles and venules, respectively, where lines denote the results of multiscale model and markers represent experimental data \cite{riva_blood_1985,werkmeister_measurement_2012,kornfield_measurement_2015}.
Black and grey lines denote the results for source distribution 1 and source distribution 2, respectively, which are almost identical.
Our model shows strong agreement with experimental data for arterioles.
The flow rates in venules are distributed more sparsely, and our model predictions lie within the observed range of these measurements.
The flow rates at CRA inlet and CRV outlet are given in table$~$\ref{flow_pressure_CRA_CRV}, with a discrepancy of $0.01~\rm{\upmu L/min}$ for both source distributions, which corresponds to approximately $0.03\%$ of CRA inlet flow rate.
Pressure differences $\Delta p$ between the entry of CRA and the individual vessel outlets are shown in figure$~$\ref{CRA_inlet_pressure_variation_flow}(c). Results for arterioles and venules are displayed in red and blue regions, respectively, with markers denoting the numerical data from other models \cite{julien_one-dimensional_2023,joldes_fundus_2016}.
In the red region, arteriole diameter decreases from left to right, whereas in the blue region, venule diameter increases from left to right.
The decrease on pressure across arteriolar tree is about $20~\rm{mmHg}$ which is greater than that across venules, at about $16~\rm{mmHg}$. 
Our model agrees well with 3D model for arteriolar diameter below $60~\rm{\upmu m}$, and with established 1D model for larger arterioles.

Figure$~$\ref{CRA_inlet_pressure_large_capillary_tissue_pressure}(a), (b) and (c) display the capillary pressure, tissue pressure and their difference for source distribution 1, respectively, while those for distribution 2 are shown in figure$~$\ref{CRA_inlet_pressure_large_capillary_tissue_pressure}(d), (e) and (f).
The red and blue circles denote the positions of arteriolar outlets and venular inlets, respectively.
The capillary pressure field varies between $32~\rm{mmHg}$ and $38~\rm{mmHg}$ for distribution 1, while for distribution 2 it varies between $30~\rm{mmHg}$ and $41~\rm{mmHg}$.
The tissue pressure field varies between $16~\rm{mmHg}$ and $21~\rm{mmHg}$ for distribution 1, while for distribution 2 it varies between $14~\rm{mmHg}$ and $25~\rm{mmHg}$.
The pressure fields for two source distributions differ substantially.
However, the capillary pressure is locally larger than the tissue pressure at each point, which results in the positive pressure difference in the entire tissue domain.
Moreover, the magnitude of tissue pressure is consistently correct, being slightly above or below $\rm{IOP}$.
Thus, our model provides a physiologically accurate description for the capillary and tissue pressure.

Table$~$\ref{flow_pressure_CRA_CRV} summarises the flow rate and pressure at CRA inlet and CRV outlet.
For both source distributions, the flow rate at the CRA inlet is nearly identical to that at the CRV outlet, with a small difference of $0.01~\rm{\upmu L/min}$.
\subsection{Pulsatile flow}
\begin{figure}
	\centering
	\includegraphics[width=0.42\textwidth]{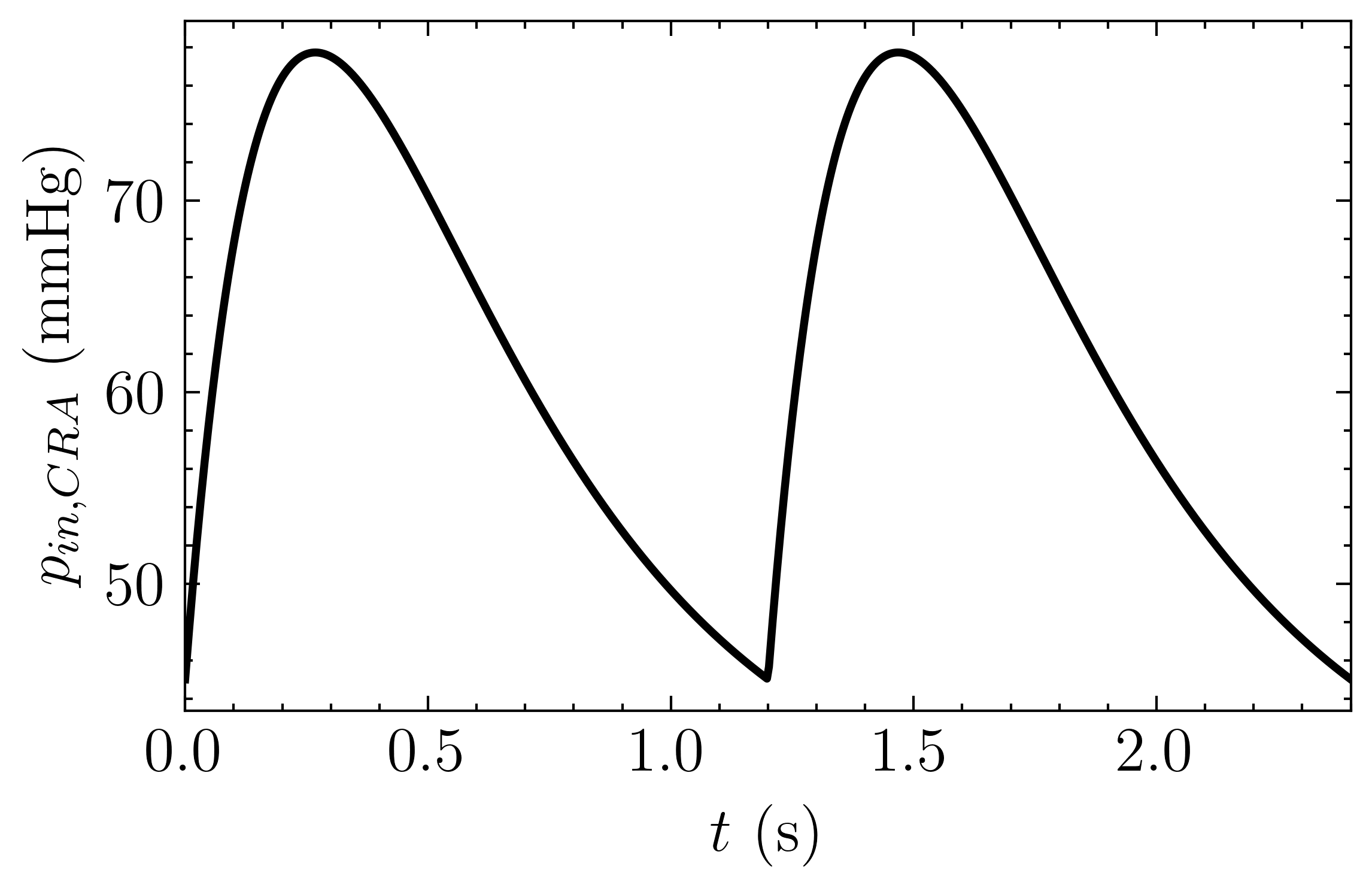}
	\caption{Pulsatile pressure profile at CRA inlet.}
	\label{p_in_CRA}
\end{figure}
The CRA inlet pressure could exhibit significant variation due to cardiac pulsation, while the CRV outlet pressure remains almost constant \cite{guidoboni_intraocular_2014}.
We next explore the retinal hemodynamics with pulsatile arteriolar pressure at CRA inlet.
A cardiac cycle period of $T=1.2$ is adopted, with peak systolic pressure at $t_p=0.268~\rm{s}$.
In the first period the variation profile is constructed using a shifted function $t\exp(-3.6t)$, whose shape is similar to a pulse profile \cite{song_multi-domain_2024}.
The waveform is scaled to produce physiologically relevant values: average pressure of $62.16~\rm{mmHg}$ approximately, with systolic and diastolic pressures of $77~\rm{mmHg}$ and $45~\rm{mmHg}$, respectively.
The pressure profile at CRA inlet is shown in figure$~$\ref{p_in_CRA}.
\begin{figure}
	\centering
	\includegraphics[width=\textwidth]{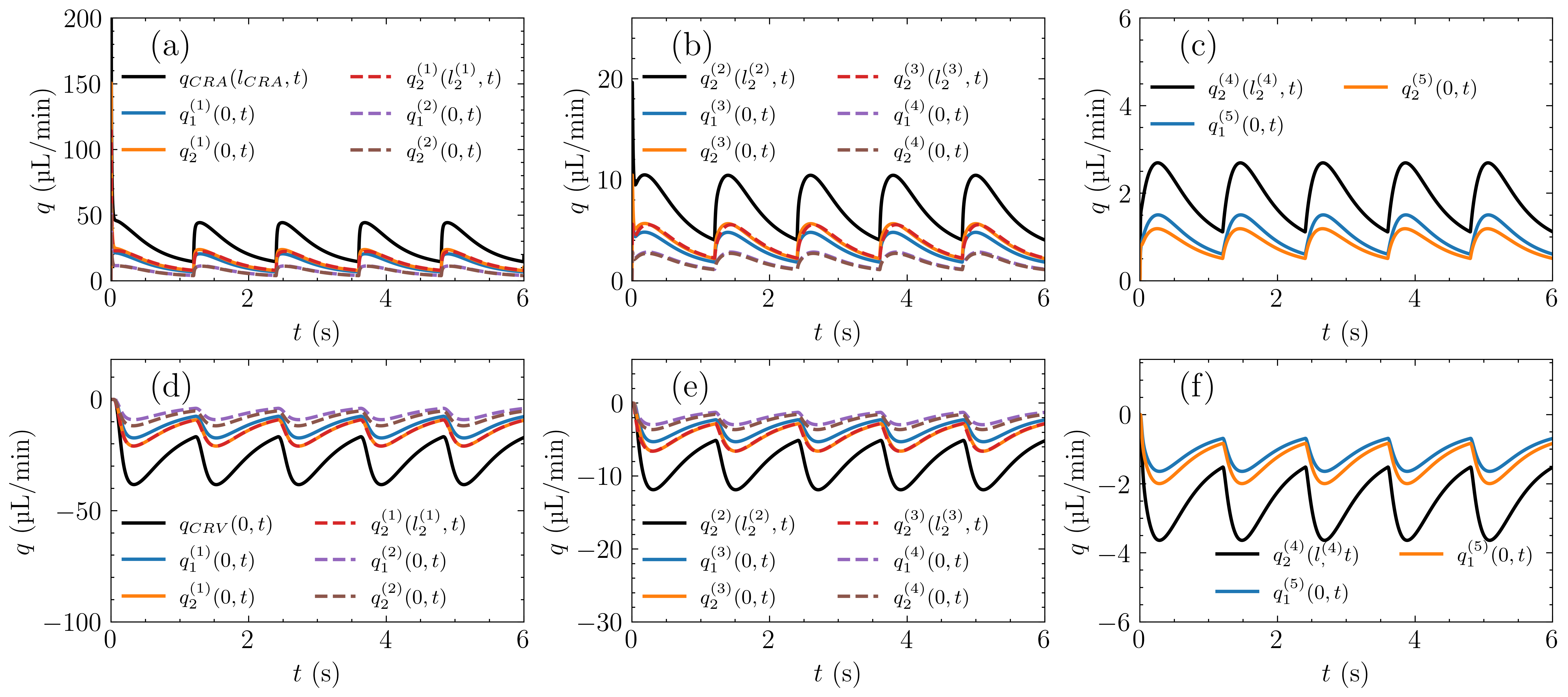}
	\caption{Temporal evolution of flow rate at bifurcation boundaries of arteriolar and venular trees. (a)-(c) Evolutions of flow rates in arteriolar tree. (d)-(f) Evolutions of flow rates in venular tree.}
	\label{pulsatile_q_arteries_veins}
\end{figure}

Figure$~$\ref{pulsatile_q_arteries_veins} shows the temporal evolution of flow rate at bifurcation boundaries in arteriolar and venular trees.
The flow rates of vessels are denoted as $q^{(n)}_k$, where $n$ is the level of bifurcation boundary and $k=1,2$ indicates the daughter vessel index, with their lengths following similar notation.
Figure$~$\ref{pulsatile_q_arteries_veins}(a), (b) and (c) show the flow through the hierarchy of arteriolar bifurcations.
In figure$~$\ref{pulsatile_q_arteries_veins}(a), solid lines show the flow rates at first-level bifurcation, including $q_{CRA}(l_{CRA},t)$ for CRA and $q^{(1)}_1(0,t)$ and $q^{(1)}_2(0,t)$ for its two daughter arterioles, while dashed lines denote the second-level flows from the secondary daughter arteriole in the first-level bifurcation.
Figure$~$\ref{pulsatile_q_arteries_veins}(b) advances to third-level flows (solid lines) and fourth-level flows (dashed lines) from the secondary branch in the upper level bifurcation.
Figure$~$\ref{pulsatile_q_arteries_veins}(c) displays flow rates at fifth-level bifurcation.
The evolution of flow rate in venules is shown in figure$~$\ref{pulsatile_q_arteries_veins}(d)-(f), following the same descending order as that for arterioles.
Flows in both CRA and CRV exhibit initial transients.
Arteriolar flows all show pulsatile oscillations similar to the CRA inlet pressure, but with different amplitudes of variations, while venular flows follow simiar pulsatility but with negative values indicating flow direction.
The variation is largest in the CRA, secondary in the CRV, and become dampened in distal arterioles and venules, with minimal oscillation at the fifth-level bifurcations.

\begin{figure}
	\centering
	\includegraphics[width=0.86\textwidth]{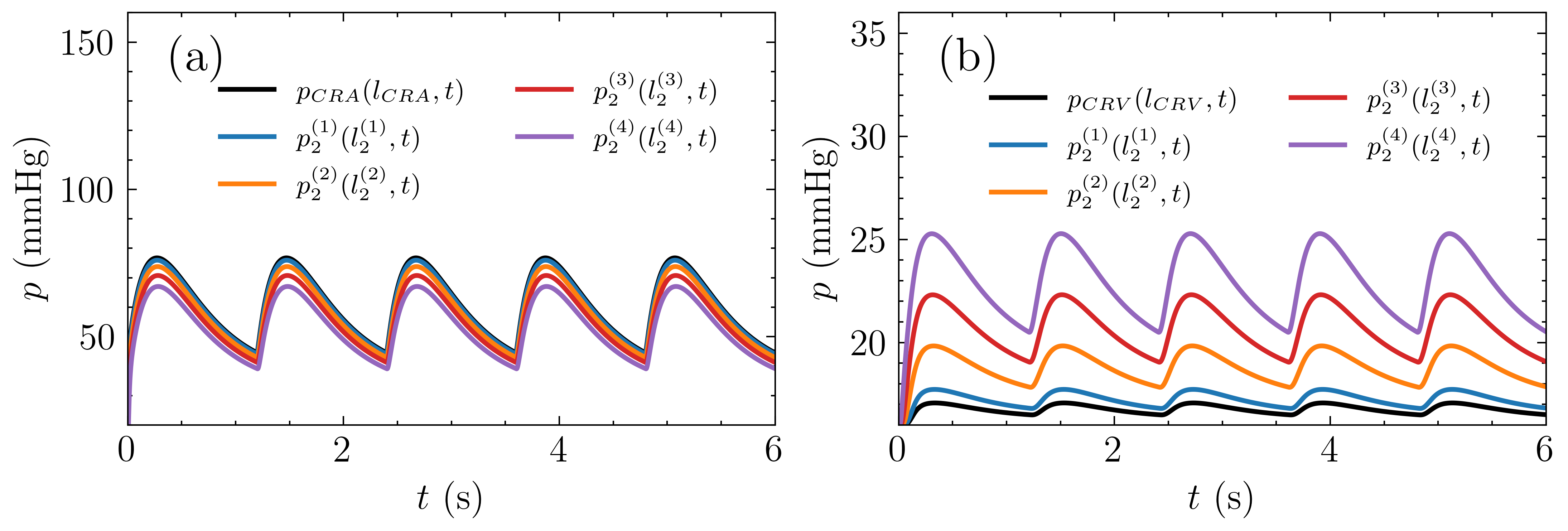}
	\caption{Evolution of pressure at bifurcation boundaries of arteriolar and venular trees. (a) Evolutions of pressures in arteriolar trees (b) Evolution of pressures in venular tree.}
	\label{pulsatile_p_arteries_veins}
\end{figure}
Figure$~$\ref{pulsatile_p_arteries_veins} shows the temporal evolution of the pressure in these vessels.
Since vessels connected to the same bifurcation boundary have identical pressure, only the pressures in mother vessels at each bifurcation are shown.
The variation of CRA is largest, and that of CRV is smallest.
The pressure waveforms in venules are more distorted than in arterioles.
Moreover the pressure amplitude is generally larger in arterioles than in venules, which results from the greater wall stiffness of arterioles.

\begin{figure}
	\centering
	\includegraphics[width=\textwidth]{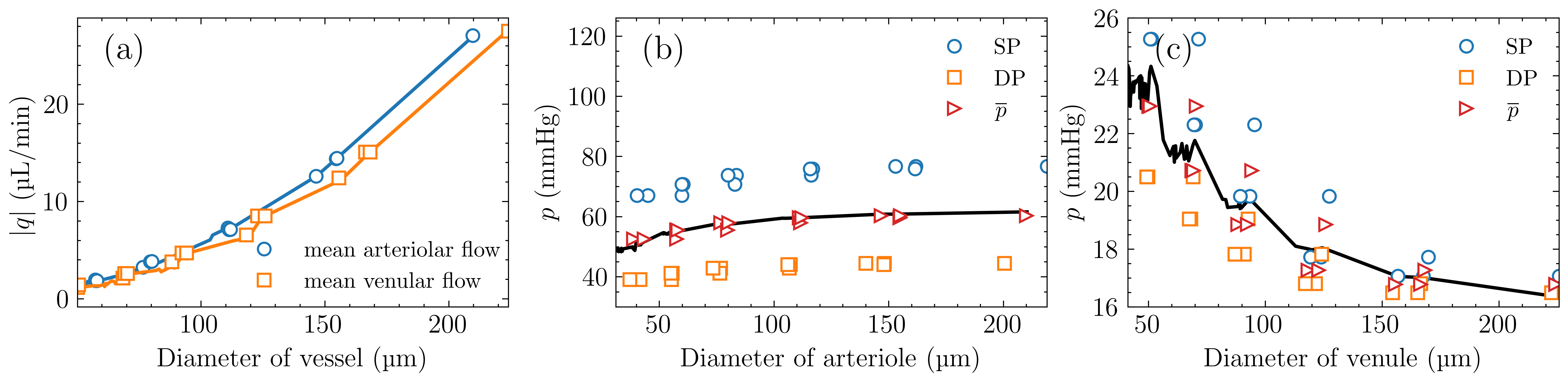}
	\caption{Pulsatile flow rates and pressures as functions of diameter. (a) Averaged pulsatile flow rates. The mean arteriolar and venular flows are denoted by blue and orange markers, respectively, while the distributions in steady flow are denoted by solid lines. (b) and (c) Arteriolar and venular pressures. The SP, DP and mean pressure are denoted by blue, orange and red markers, respectively, and the pressure distribution with constant CRA inlet pressure is denoted by black line.}
	\label{pulsatile_flow_mean_arteries_veins}
\end{figure}
The flow rates and pressures shown in figures$~$\ref{pulsatile_q_arteries_veins} and \ref{pulsatile_p_arteries_veins} are averaged and compared to those in steady flow.
Figure$~$\ref{pulsatile_flow_mean_arteries_veins}(a) shows the mean arteriolar and venular flow rates, denoted by blue and orange markers, respectively.
The flow rate distributions for arterioles and venules in steady flow are displayed in blue and orange lines, respectively.
The mean arteriolar and venular flows under pulsatile conditions closely match the solutions in steady flow, which results from the longer diastolic phase as well as the time-averaged CRA inlet pressure that is nearly identical to the constant CRA inlet pressure in steady flow.
Figure$~$\ref{pulsatile_flow_mean_arteries_veins}(b) and (c) show the systolic pressure (SP), diastolic pressure (DP) and mean pressure $\overline{p}$ as functions of the diameter in arterioles and venules, respectively.
The SP, DP and mean pressure are denoted by blue, orange and red markers, respectively.
As a baseline, the pressure distribution under constant CRA inlet pressure $p_{in,CRA}=62~\rm{mmHg}$ is shown by black line.
In steady and pulsatile flows all arteriolar pressures generally increase with diameter but non-monotonically, which indicates the dependence of pressures not only on the diameter but also on the distance from the CRA inlet \cite{julien_one-dimensional_2023}.
Venular pressures decrease with diameter and the decreasing magnitude is smaller than the increasing magnitude in arterioles.
Moreover, both arteriolar and venular pressures align closely with corresponding steady distributions.

Table$~$\ref{flow_pressure_CRA_CRV} summarises the averaged flow rate and pressure at CRA inlet and CRV outlet.
The magnitudes of averaged flow rates at both CRA inlet and CRV outlet are slightly larger than in steady condition due to the higher time-averaged CRA inlet pressure.
The flow rates at CRA inlet and CRV outlet are almost identical, thus the conservation of mass is also satisfied. 
\begin{table}
	\centering
	\begin{tabular}{l l l l}
		& $q_{in,CRA}$ & $p_{in,CRA}$ & $q_{out,CRV}$\\
		\hline
		Steady flow (distribution 1) & $27.318~\rm{\upmu L/min}$ & $62~\rm{mmHg}$ & $-27.330~\rm{\upmu L/min}$ \\
        Steady flow (distribution 2) & $26.261~\rm{\upmu L/min}$ & $62~\rm{mmHg}$ & $-26.273~\rm{\upmu L/min}$ \\
		Pulsatile flow &$27.338~\rm{\upmu L/min}$ & $62.159~\rm{mmHg}$ & $-27.797~\rm{\upmu L/min}$ \\
	\end{tabular}
	\caption{The flow rates and pressures at CRV outlet and CRA inlet.}
	\label{flow_pressure_CRA_CRV}
\end{table}

\section{Sensitivity analysis of parameters in coupled system}
\label{SensitivityAnalysis}
We next perform sensitivity analysis of the crucial parameters in capillary-tissue coupled system, including capillary permeability $k_{cap}$, tissue permeability $k_t$ and exchange rate $\alpha_{exch}$.
To quantify the sensitivity, we use the CRA inlet flow rate $q_{in,CRA}$ as an indicator for vascular trees and the domain-averaged pressures $\langle p_{cap}\rangle$ and $\langle p_t\rangle$ as indicators for the capillary-tissue system, given by equations$~$(\ref{p_cap_domain_average}) and (\ref{p_t_domain_average}).
The oncotic pressure difference is not explored in this section, since the dependence of these variables on it is simple: the CRA inlet flow rate is insensitive to it, as rationalised in section$~$\ref{multiscale_coupling_condition}, and it only adds a constant offset to capillary and tissue pressures.
The results of $q_{in,CRA}$, $\langle p_{cap}\rangle$ and $\langle p_t\rangle$ for baseline values are denoted by $q^0_{in,CRA}$, $\langle p^0_{cap}\rangle$ and $\langle p^0_t\rangle$, respectively, which are summarised in table$~$\ref{sensitivity_baseline_values}.

For each parameter analysed, the other two parameters are fixed at their baseline values.
Since $k_{cap}$, $k_t$ and $\alpha_{exch}$ are effective parameters and direct experimental data precisely related to them are scarce, we adopt a prudent approach by allowing these parameters to vary over several orders of magnitude, which correspond to a $99\%$ decrease and a $9900~\%$ increase, respectively, relative to the baseline value.
This enables a comprehensive sensitivity analysis that spans the full range of plausible values and identifies the regimes where they affect the retinal hemodynamics significantly.
The baseline values for these three parameters and the rationale for them are provided in Appendix$~$\ref{baseline_coupled_system}.
The remaining parameters, including those in the 1D model, viscosities $\mu_{cap}$ and $\mu_t$ and the radius of tissue domain $R_t$, are summarised in table$~$\ref{parameters_retinal_circulation}.
\subsection{Sensitivity of capillary permeability}
The effective permeability of capillary bed $k_{cap}$ governs the resistance to flow through the capillaries.
We perform a sensitivity analysis over a plausible range to identify the regimes where $k_{cap}$ could significantly affect the retinal hemodynamics.
The baseline value is $k^0_{cap}=0.03~\rm{\upmu m}^2$.
\begin{figure}
    \centering
    \includegraphics[width=\textwidth]{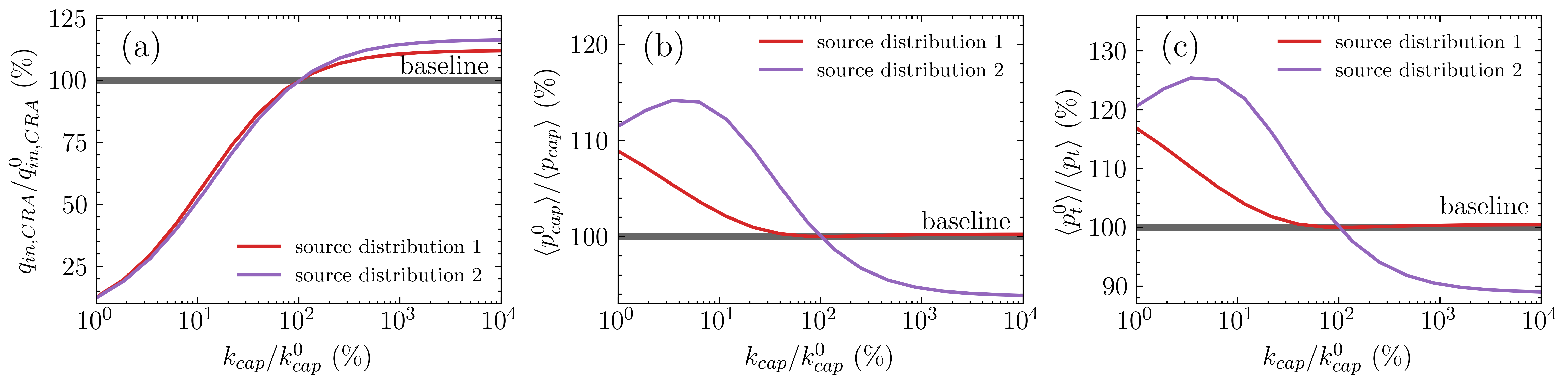}
    \caption{Sensitivity of CRA inlet flow rate and domain-averaged pressures to capillary permeability. (a) Sensitivity of CRA inlet flow rate to capillary permeability. (b) Sensitivity of domain-averaged capillary pressure to capillary permeability. (c) Sensitivity of domain-averaged tissue pressure to capillary permeability.}
    \label{sensitivity_capillary_permeability}
\end{figure}

Figure$~$\ref{sensitivity_capillary_permeability}(a) illustrates the sensitivity of CRA inlet flow rate $q_{in,CRA}$ to capillary permeability, where the results for distribution 1 and distribution 2 are denoted by red and purple lines, respectively.
For both source distributions, $q_{in,CRA}$ is asymmetrically sensitive to capillary permeability.
Decreasing $k_{cap}$ to $1\%$ of its baseline value decreases flow by about $90\%$, whereas increasing $k_{cap}$ to $100$ times the baseline increases flow by only about $20\%$.

Figure$~$\ref{sensitivity_capillary_permeability}(b) and (c) display the sensitivity of domain-averaged capillary pressure $\langle p_{cap}\rangle$ and domain-averaged tissue pressure $\langle p_t\rangle$ to capillary permeability, respectively.
For distribution 1, both averaged pressures are higher than their baseline values when $k_{cap}$ is below its baseline, and they increase further as $k_{cap}$ decreases.
Above the baseline, pressures saturate around their baseline values.
Similarly, for distribution 2, the averaged pressures are higher than their baselines when $k_{cap}$ is below its baseline, but exhibit a non-monotonic dependence on $k_{cap}$.
Above the baseline, $\langle p_{cap}\rangle$ and $\langle p_t\rangle$ tends to about $92\%$ and $90\%$ of their baseline values, respectively.

These results show that capillary permeability can significantly affect $q_{in,CRA}$, while the variations for $\langle p_{cap}\rangle$ and $\langle p_t\rangle$ across all values of $k_{cap}$ are more modest.
Moreover, while capillary and tissue pressures largely depend on the specific source distribution, the flow in arteriolar and venular trees is less sensitive to it.
\begin{table}
	\centering
	\begin{tabular}{l l l l l}
		& $q^0_{in,CRA}$ & $\langle p^0_{cap}\rangle$ & $\langle p^0_t\rangle$ \\
		\hline
		Source distribution 1 & $27.318~\rm{\upmu L/min}$ & $33.847~\rm{mmHg}$ & $17.847~\rm{mmHg}$\\
        Source distribution 2 & $26.261~\rm{\upmu L/min}$ & $36.174~\rm{mmHg}$ & $20.174~\rm{mmHg}$ &
	\end{tabular}
	\caption{Baseline values of variables in sensitivity analysis.}
	\label{sensitivity_baseline_values}
\end{table}

\subsection{Sensitivity of tissue permeability}
Existing experimental data reveal considerable variation in retinal hydraulic conductivity, spanning two orders of magnitude difference \cite{fatt_flow_1971,antcliff_hydraulic_2001,ruffini_mathematical_2024}.
Understanding how this variability affects retinal hemodynamics is thus intriguing.
The baseline value is $k^0_t=0.0002~\rm{\upmu m}^2$.
\begin{figure}
    \centering
    \includegraphics[width=\textwidth]{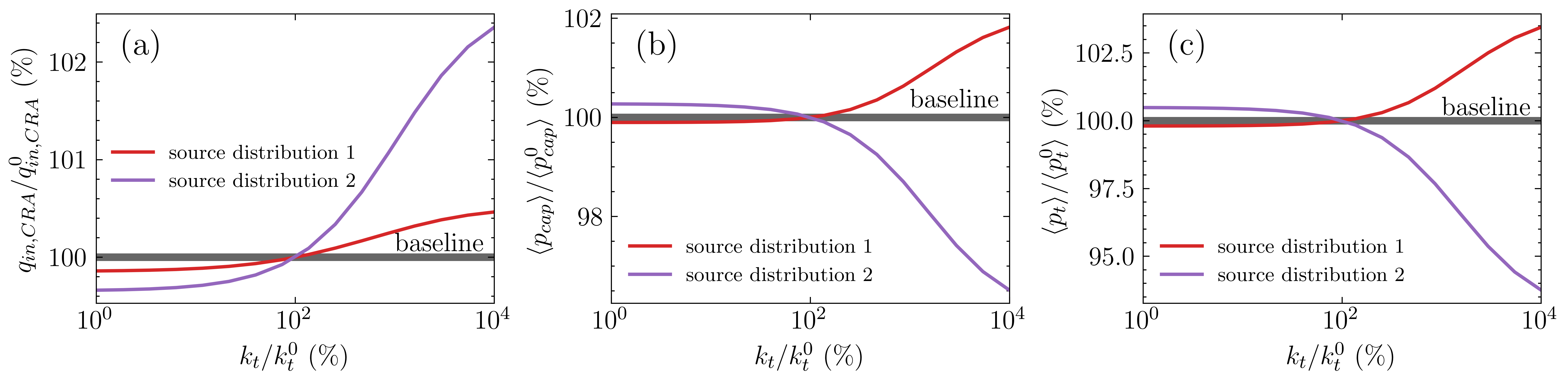}
    \caption{Sensitivity of CRA inlet flow rate and domain-averaged pressures to tissue permeability. (a) Sensitivity of CRA inlet flow rate. (b) Sensitivity of domain-averaged capillary pressure. (c) Sensitivity of domain-averaged tissue pressure.}
    \label{sensitivity_tissue_permeability}
\end{figure}

Figure$~$\ref{sensitivity_tissue_permeability}(a), (b) and (c) illustrate the sensitivity of CRA inlet flow rate $q_{in,CRA}$, averaged capillary pressure $\langle p_{cap}\rangle$ and averaged tissue pressure $\langle p_t\rangle$ to tissue permeability, respectively.
The saturation for all three variables occurs below the baseline, with the saturation value at approximately $k_t=0.0001~\rm{\upmu m}^2$.
This might suggest that the pathologies affecting tissue permeability would only impair fluid exchange if $k_t$ is below the baseline.

The magnitude of variation for $q_{in,CRA}$ across the full range of $k_t$ is less than $3\%$ for both source distributions.
Moreover, $q_{in,CRA}$ increases as $k_t$ increases for both source distributions.
The domain-averaged pressures for the two source distributions exhibit opposite dependencies on $k_t$: while the pressures for source distribution 1 increase as $k_t$ increases, those for source distribution 2 decrease as $k_t$ increases.
The magnitude of variation for both pressures is about $2\%$.
Thus all variables are less sensitive to $k_t$ than to $k_{cap}$. 
This minimal sensitivity of flow and pressures to tissue permeability indicates that tissue permeability does not play a crucial role in retinal hemodynamics.
\subsection{Sensitivity of exchange rate}
In our model, the exchange rate $\alpha_{exch}=L_h\times S$ reflects the combined effect of capillary wall hydraulic conductivity $L$ and exchange surface area $S$.
An increase in $L_h$ could be related to BRB dysfunction, while a change in $S$ could arise from capillary dilation or pathological angiogenesis.
\begin{figure}
    \centering
    \includegraphics[width=\textwidth]{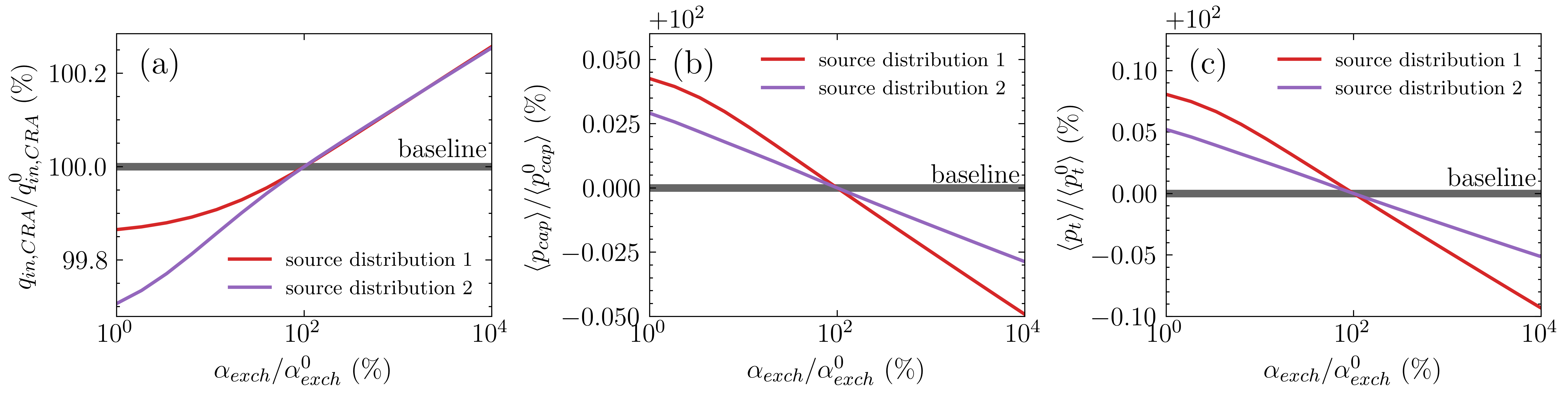}
    \caption{Sensitivity of CRA inlet flow rate and domain-averaged pressures to exchange rate. (a) Sensitivity of CRA inlet flow rate. (b) Sensitivity of domain-averaged capillary pressure to exchange rate. (c) Sensitivity of domain-averaged tissue pressure to exchange rate.}
    \label{sensitivity_exchange}
\end{figure}

Figure$~$\ref{sensitivity_exchange}(a), (b) and (c) show the sensitivity of $q_{in,CRA}$, $\langle p_{cap}\rangle$ and $\langle p_t\rangle$ to $\alpha_{exch}$, respectively.
All three variables are largely insensitive to exchange rate, with very small variations.
Specifically, the variation in $q_{in,CRA}$ is the largest at about $0.2\%$.
For both source distributions $q_{in,CRA}$ increases as $\alpha_{exch}$ increases, and interestingly, the two curves converge to the same value above the baseline $\alpha^0_{exch}$, showing that $q_{in,CRA}$ is independent of the specific distribution of source points above the baseline $\alpha^0_{exch}$.
The domain-averaged pressures decrease as $\alpha_{exch}$ decreases.
Such insensitivity suggests that although $\alpha_{exch}$ governs local fluid exchange, it has negligible effects on the global hemodynamic variables $q_{in,CRA}$, $\langle p_{cap}\rangle$ and $\langle p_t\rangle$.
\section{Discussion and conclusion}
\label{DiscussionAndConclusion}
This work proposes a multiscale computational model for modelling hemodynamics in retinal microcirculation, by extending previous work on tree-type vasculature flow model for arterioles and venules and porous-medium-type model for capillaries and tissue.
We have developed a fully coupled flow model that links arterioles, venules, capillaries and interstitial tissue in a unified framework.
In particular, the 1D model \cite{julien_one-dimensional_2023} is adopted for the flow of arteriolar and venular trees in the superficial layer, while vessels in the deep layer are modelled as effective resistors.
The capillary bed and surrounding tissue are treated as interacting porous media, which are modelled by coupled Darcy models.
Our approach is based on widely-accepted physical principles, cf. \cite{eisenberg_structural_2023}, with necessary simplifications.

Another key novelty of our approach is the derivation and implementation of the analytic solution for capillary-tissue coupled system, which extends previous works that use the analytic solution for capillary flow in multiscale modelling \cite{shipley_hybrid_2019,sweeney_threedimensional_2024,desposito_computational_2018}.
The capillary-tissue coupled system is solved analytically by using a decoupling transformation, which separates governing equations into two components, i.e., mean pressure and exchange pressure.
The solutions for capillary and tissue pressures are expressed as the combinations of logarithmic functions and infinite series of modified Bessel functions.
The convergence and truncation error of infinite series solution are analysed extensively to ensure the robustness and accuracy of the model.
This also suggests a potential modelling methodology for other tissues with complicated geometries, where the boundary-sensitive mean pressure could be solved numerically while the exchange pressure, which controls local perfusion, can be effectively approximated using a free-space solution.
Furthermore, we establish a dynamic coupling condition that bridges the capillary bed with upstream arteriolar and downstream venular flows.
This multiscale coupling, including integral term and nonlinear pressure-area relation, is then simplified for efficient computation.

Our model's predictions with a constant baseline CRA inlet pressure agree well with experimental data, demonstrating its high predictive accuracy.
We also simulate the pulsatile flows with a cardiac-cycle pressure profile at CRA inlet.
The results show that the model can characterise pulsations of the pressure and flow in arterioles and venules.
Moreover, the sensitivity analysis of the parameters in capillary-tissue system is performed. 
The hemodynamics is most sensitive to capillary permeability, followed by tissue permeability and exchange rate, and possible physiological implications are discussed.
The flow depends on specific distribution of source points, and further exploration of how different distributions of source points affect retinal hemodynamics would be valuable.
It is also worth noting that saturation occurs when permeabilities move below or above certain values, which might correspond to the lower or upper bound of physiological conditions.

Despite its effectiveness, our model has several limitations.
First, we have assumed constant permeabilities for the capillaries and tissue for simplicity, even though we are aware
that biology and evolution are likely to use spatially variable permeabilities. Indeed, the type of permeability might be expected to vary from location to location reflecting the specialized distribution of ion channels and transporters used by evolution to create overall function.
Work is currently underway to extend our work to spatially varying permeabilities which will allow us to explore physiologically relevant questions such as the effect of non-uniform nature of the capillaries. We are also extending our model to time-varying permeability of the capillaries to explore the effect of neurovascular coupling.      
Second, we approximate the area where the retinal vasculature is distributed as a two-dimensional circular domain, and further studies could consider a three-dimensional spherical geometry which would represent the actual retinal anatomy better.
Additionally, the model does not distinguish different capillary plexuses located in outer retina and inner retina \cite{causin_blood_2016}, and this simplification could be addressed.
Notably, several works have developed transport models for water and various organic and inorganic molecules in the optic nerve \cite{song_multi-domain_2024,zhu_bidomain_2019,zhu_membranes_2021,xiao_potassium_2025,xiao_glymphatic_2025,zhu_tridomain_2021,zhu_optic_2021}, and several fundamental properties of the transport in optic nerve have been revealed.
Combining our multiscale model with such models could provide deeper insights into retinal physiology and pathology.
Similarly, adding the active pumping of retinal pigment epithelium and passive exchange with vitreous chamber will help to describe more physiologically relevant situations.
Finally, the boundary condition for tissue pressure enforces no fluid accumulation in the tissue, and it might not be physiologically accurate in certain pathological situations.
For example, exudative macular edema is caused by the fluid accumulation in the tissue \cite{ruffini_mathematical_2024}.
Further studies could thus explore other boundary conditions for such pathological situations.

In conclusion, the multiscale model proposed in this work provides an efficient and robust tool for exploring the hemodynamics in retinal microcirculation.
Despite the necessary simplifications on which the model is based, our model generates physiologically consistent results and could thus help to understand the mechanisms of retinopathies as well as perform the computation for specific patients.
\vskip 12pt

\noindent \textbf{Acknowledgement.} This work is supported in part by NSFC (Project number 12231004). We thank the reviewers whose comments motivated significant improvements to the paper.

\noindent \textbf{Declaration of interests.} The authors report no conflict of interest.

\appendix
\newpage

\section{Solution for mean pressure}
\label{SolutionMeanPressure}
Enforcing the boundary condition $\nabla^\prime f_{mean}({\bm x},{\bm x}^\prime)\cdot {\bm n}^\prime=2\pi R_t$ yields to
\begin{eqnarray}
    \frac{1}{\vert{\bm x}^\prime-{\bm x}\vert^2}\left[\vert {\bm x}^\prime\vert^2+k\vert {\bm x}\vert^2-(k+1){\bm x}^\prime\cdot{\bm x}\right]=1,
\end{eqnarray}
which holds for all ${\bm x}$ and ${\bm x}^\prime$ if and only if $k=1$, thus the coefficient $k$ is uniquely determined.
The constant $c$ remains arbitrary and can be set as $c=0$ without loss of generality.
Thus the solution to the adjoint problem is
\begin{eqnarray}
    f_{mean}({\bm x},{\bm x}^\prime)=\frac{1}{2\pi}\left(\log\vert{\bm x}^\prime-{\bm x}\vert+\log \vert{\bm x}^\prime-{\bm x}^\dprime\vert\right).
\end{eqnarray}
Then the solution for $p_{mean}$ can obtained via following convolution \cite{riley_mathematical_2006,kim_proof_1993}
\begin{eqnarray}
    p_{mean}=\int_{\Omega_t}f_{mean}{\nabla^\prime}^2 p_{mean}({\bm x}^\prime)d{\bm x}^\prime+\int_{\partial \Omega_t}\left[p_{mean}({\bm x}^\prime)\nabla^\prime f_{mean}-f_{mean}\nabla^\prime p_{mean}({\bm x}^\prime)\right]\cdot {\bm n}^\prime ds^\prime.
\end{eqnarray}
Substituting the boundary conditions for $f_{mean}$ and $p_{mean}$ yields to
\begin{eqnarray}
    p_{mean}({\bm x})=\int_{\Omega_t}f_{mean}({\bm x},{\bm x}^\prime){\nabla^\prime}^2 p_{mean}({\bm x}^\prime)d{\bm x}^\prime+\frac{1}{S}\int_{\partial \Omega_t}p_{mean}({\bm x}^\prime) ds^\prime.
\end{eqnarray}
The first integral is computed as
\begin{eqnarray}
    \int_{\Omega_t}f_{mean}({\bm x},{\bm x}^\prime){\nabla^\prime}^2 p_{mean}({\bm x}^\prime)d{\bm x}^\prime&=&-\frac{1}{\kappa_{cap}+\kappa_t}\sum^{N_{cap}}_{n=1}q_n\int_{\Omega_t}f_{mean}({\bm x},{\bm x}^\prime)\delta({\bm x}^\prime-{\bm x}_n)d{\bm x}^\prime \nonumber\\
    &=&-\frac{1}{\kappa_{cap}+\kappa_t}\sum^{N_{cap}}_{n=1}q_nf_{mean}({\bm x},{\bm x}_n).
\end{eqnarray}

\section{Solution for the Green's function of exchange pressure}
\label{SolutionExchangePressure}
The continuity of the function $f_{exch}$ at $r=r_n$ provides that
\begin{eqnarray}
    R_{m,n}(r)\vert_{r=r^+_n}-R_{m,n}(r)\vert_{r=r^-_n}=0,
    \label{continuity_radial_func}
\end{eqnarray}
and the discontinuity of the derivative of $f_{exch}$ implies that
\begin{eqnarray}
    \left.\frac{dR_{m,n}(r)}{dr}\right\vert_{r=r^+_n}-\left.\frac{dR_{m,n}(r)}{dr}\right\vert_{r=r^-_n}=\frac{1}{r_n}.
    \label{discontinuity_radial_function}
\end{eqnarray}
Moreover imposing the no flux boundary condition yields
\begin{eqnarray}
    \left.\frac{dR_{m,n}(r)}{dr}\right\vert_{r=R_t}=0.
    \label{flux_radial_function}
\end{eqnarray}
The boundary condition $dR_{m,n}(r)/dr\vert_{r=R_t}=0$ provides that
\begin{eqnarray}
    C_{m,n}=-\frac{I^\prime_m(\lambda R_t)}{K^\prime_m(\lambda R_t)}B_m,
\end{eqnarray}
and the continuity of $R_m(r)$ yields
\begin{eqnarray}
    A_{m,n}=\left[1-\frac{I^\prime_m(\lambda R_t)}{K^\prime_m(\lambda R_t)}\frac{K_m(\lambda r_n)}{I_m(\lambda r_n)}\right]B_m.
\end{eqnarray}
Thus the discontinuity of $dR_m(r)/dr$ at $r=r_n$ gives rise to
\begin{eqnarray}
    B_{m,n}\frac{I^\prime_m(\lambda R_t)}{K^\prime_m(\lambda R_t)}\frac{I_m(\lambda r_n)K^\prime_m(\lambda r_n)-K_m(\lambda r_n)I^\prime_m(\lambda r_n)}{I_m(\lambda r_n)}=\frac{1}{\lambda r_n},
\end{eqnarray}
and using the identity \cite[p. 866]{baricz_bounds_2016}
\begin{eqnarray}
    I_m(\lambda r_n)K^\prime_m(\lambda r_n)-K_m(\lambda r_n)I^\prime_m(\lambda r_n)=-\frac{1}{\lambda r_n}
\end{eqnarray}
we obtain
\begin{eqnarray}
    B_{m,n}=\frac{K^\prime_m(\lambda R_t)}{I^\prime_m(\lambda R_t)}I_m(\lambda r_n),
\end{eqnarray}
and
\begin{eqnarray}
    A_{m,n}&=&\frac{K^\prime_m(\lambda R_t)I_m(\lambda r_n)}{I^\prime_m(\lambda R_t)}-K_m(\lambda r_n),\\
    C_{m,n}&=&-I_m(\lambda r_n).
\end{eqnarray}

\section{Ratio between terms}
\label{RadtioRadialComponent}
The ratio between the first term and the second term in radial component is given by
\begin{eqnarray}
    e_{m,n}=\left\vert \frac{K^\prime_m(\lambda R_t)I_m(\lambda r^n_>)}{I^\prime_m(\lambda R_t)K_m(\lambda r^n_>)}\right\vert.
\end{eqnarray}
If $m$ is fixed the derivative $K^\prime_m(\lambda r)$ is negative and increasing.
Then
\begin{eqnarray}
    e_{m,n} \leq -\frac{K^\prime_m(\lambda R_t)I_m(\lambda r^n_>)}{I^\prime_m(\lambda r^n_>)K_m(\lambda r^n_>)}.
\end{eqnarray}
Using the inequality that \cite[p. 867]{baricz_bounds_2016}
\begin{eqnarray}
    \frac{I^\prime_m(\lambda r)}{I_m(\lambda r)}>\frac{m}{\lambda r},\quad m>-1, \quad \lambda r>0,
\end{eqnarray}
the inequality can be written as
\begin{eqnarray}
    e_{m,n} \leq -\frac{\lambda r^n_>}{m}\frac{K^\prime_m(\lambda R_t)}{K_m(\lambda r^n_>)}, \quad m>-1, \quad \lambda r>0.
\end{eqnarray}
Using the inequality that \cite[p. 867]{baricz_bounds_2016}
\begin{eqnarray}
    -\frac{K^\prime_m(\lambda r)}{K_m(\lambda r)}<\frac{m}{\lambda r}+1, \quad m>1/2,\quad \lambda r>0,
\end{eqnarray}
we obtain
\begin{eqnarray}
    e_{m,n} \leq \frac{\lambda r^n_>}{m}\left(\frac{m}{\lambda R_t}+1\right)\frac{K_m(\lambda R_t)}{K_m(\lambda r^n_>)}, \quad m>1/2,\quad \lambda r>0.
\end{eqnarray}
Using the inequality (\citealp[equation$~$3.4]{baricz_bounds_2010}; \citealp[equation$~$2.17]{joshi_inequalities_1996}) that $K_m(r_1)/K_m(r_2)>(r_2/r_1)^m$ for $m>-1/2$ and $0<r_1<r_2$ we obtain that
\begin{eqnarray}
    e_{m,n} \leq \frac{\lambda r^n_>}{m}\left(\frac{m}{\lambda R_t}+1\right)\left(\frac{\lambda r^n_>}{\lambda R_t}\right)^m, \quad m>1/2,\quad \lambda r>0,
\end{eqnarray}
thus for $0\leq r<R_t$ the ratio $e_{m,n}\to 0$ as $m\to\infty$ and the first term can be neglected.

\section{Proof of addition formula for infinite series}
\label{ProofGrafAddition}
Consider following problem in two dimensions:
\begin{eqnarray}
	\nabla^2f_{exch}({\bm x},{\bm x}_n)-\lambda^2 f_{exch}({\bm x},{\bm x}_n)=\delta({\bm x}-{\bm x}_n),
\end{eqnarray}
subjected to the boundary condition that $f_{exch}\to 0$ as $\vert {\bm x}\vert\to \infty$.
In polar coordinates, with ${\bm x}_n=(r_n,\theta_n)$, the equation is given by equation$~$(\ref{radial_angle}).
Expanding $f_{exch}$ in a Fourier series and solving the resulting modified Bessel equation, with proper matching for the coefficients of the expansion of delta function, yields
\begin{eqnarray}
	f_{exch}=-\frac{1}{2\pi}\sum^{\infty}_{m=-\infty}I_m(\lambda r_<)K_m(\lambda r_>)\exp[im(\theta-\theta_n)],
\end{eqnarray}
and $r^n_<=\min(r,r_n)$ and $r^n_>=\max(r_,r_n)$.
On the other hand, since the problem is defined on infinite domain $f_{exch}$ is independent of angular variable.
The equation becomes
\begin{eqnarray}
	\frac{1}{r}\frac{\partial}{\partial r}\left(r\frac{\partial f_{exch}}{\partial r}\right)-\lambda^2f_{exch}=\frac{\delta(r-r_n)\delta(\theta-\theta_n)}{r},
\end{eqnarray}
and solving this equation with given boundary condition we obtain
\begin{eqnarray}
	f_{exch}=-\frac{1}{2\pi}K_0\left(\lambda\sqrt{r^2+r^2_n-2rr_n\cos(\theta-\theta_n)}\right).
\end{eqnarray}
Then the uniqueness of the solution implies the addition formula for modified Bessel functions.

\section{Range of parameters in coupled system}
\label{baseline_coupled_system}
The effective permeability of the capillary bed in porous medium models varies across different tissues.
In previous work on retinal hemodynamics, a hydraulic conductivity of $0.1~\rm{mm^2/(s\cdot kPa)}$ was used in the Darcy model for retinal capillaries \cite{dziubek_effect_2016}, which yields $k_{cap}\approx 0.3~\rm{\upmu m}^2$ using a blood viscosity $\mu_{cap}\approx 3~\rm{cP}$.
We adopt a moderate capillary permeability $k^0_{cap}=0.03~\rm{\upmu m}^2$ for baseline value and consider a plausible range from $k_{cap}=0.0003~\rm{\upmu m}^2$ to $k_{cap}=3~\rm{\upmu m}^2$.

For the interstitial fluid in tissue, clinical observations provide sparse values with even two orders of magnitude difference.
The hydraulic conductivity in human retina is $2.54\times 10^{-10}~\rm{m/s/Pa}$ \cite{ruffini_mathematical_2024,antcliff_hydraulic_2001}, which corresponds to an intrinsic permeability of approximately $6\times10^{-5}~\rm{\upmu m}^2$ with a fluid viscosity $\mu_t=0.7~\rm{cP}$.
Moreover the range of tissue permeabilities extends to approximately on the order of $10^{-2}~\rm{\upmu m}^2$ \cite{fatt_flow_1971}.
Given this variation, the baseline tissue permeability is then set as $k^0_t=0.0002~\rm{\upmu m}^2$, with a plausible range from $2\times10^{-5}~\rm{\upmu m}^2$ to $0.002~\rm{\upmu m}^2$.

The exchange rate between capillaries and tissue $\alpha_{exch}$ depends on capillary wall hydraulic conductivity $L_h$ and the exchange surface area per unit tissue volume $S$.
Given that retinal capillaries are fundamentally similar in barrier structure to those in brain \cite{bradbury_blood-brain_1990}, we use the value of hydraulic conductivity $L_h=0.2\times 10^{-7}~\rm{cm/s/cm~H_2O^{-1}}$ from clinical observations for human brain \cite{helton_novel_2017}.
Considering the characteristic capillary radius $6~\rm{\upmu m}$ and the volume fraction $1\%$, then $S\approx 30~\rm{cm}^{-1}$ which yields a value $\alpha_{exch}\approx 6\times10^{-8}~\rm{cm \cdot s/g}$.
We then adopt a moderate value $\alpha^0_{exch}=0.0002~\rm{\upmu m \cdot s/g}$ for baseline value and a plausible range of $\alpha_{exch}$ from $\alpha_{exch}=2\times10^{-6}~\rm{\upmu m \cdot s/g}$ to $\alpha_{exch}=0.02~\rm{\upmu m \cdot s/g}$.

\bibliographystyle{elsarticle-num.bst}
\bibliography{Ref.bib}

\end{document}